\documentclass[11pt,a4paper]{article}
\pdfoutput=1

\usepackage[pdftex]{graphics}
\usepackage{jheppub}
\usepackage{graphicx}
\usepackage{slashed}
\newcommand{\be}{\begin{equation}}
\newcommand{\ee}{\end{equation}}
\newcommand{\beq}{\begin{equation}}
\newcommand{\eeq}{\end{equation}}
\newcommand{\bea}{\begin{eqnarray}}
\newcommand{\eea}{\end{eqnarray}}

\newcommand{\gsim}{\lower.7ex\hbox{$\;\stackrel{\textstyle>}{\sim}\;$}}
\newcommand{\lsim}{\lower.7ex\hbox{$\;\stackrel{\textstyle<}{\sim}\;$}}

\newcommand{\llsim}{\lower.7ex\hbox{$\;\stackrel{\textstyle<}{\sim}\;$}}
\newcommand\eg{{\it {e.g.}}}

\newcommand\ie{{\it i.e.}}

\newcommand{\BR}{{\rm BR}}

\title{\boldmath Resonant Higgs boson pair production in the
  $hh\to b\bar{b} \; WW \to b\bar{b} \ell^+ \nu \ell^-
  \bar\nu$ decay channel}
\author[a,b]{V\'ictor Mart\'in Lozano,}
\author[a]{Jes\'us M. Moreno}
\author[c,d]{and Chan Beom Park}
\affiliation[a]{Instituto de F\'{\i}sica Te\'orica, IFT-UAM/CSIC,
  Nicol\'as Cabrera 13,\\
  UAM Cantoblanco, 28049 Madrid, Spain.}
\affiliation[b]{Dept. F\'{\i}sica  Te\'orica, Universidad Aut\'onoma de Madrid,  28049 Madrid, Spain.}
\affiliation[c]{Korea Institute for Advanced Study, Seoul 130--722, Korea.}
\affiliation[d]{CERN, Theory Division, 1211 Geneva 23, Switzerland.}
\emailAdd{victor.martinlozano@uam.es}
\emailAdd{jesus.moreno@csic.es}
\emailAdd{cbpark@kias.re.kr}

\abstract{
  Adding a scalar singlet provides one of the simplest extensions of
  the Standard Model. In this work we briefly review the latest
  constraints on the mass and mixing of the new Higgs boson and study
 its production and decay at the LHC.
  We mainly focus on double Higgs production in the
  $hh \to b \bar{b} WW \to  b \bar{b} \ell^+ \nu \ell^- \bar{\nu}$
  decay channel. This decay is found to be efficient in a region of
  masses of the heavy Higgs boson of 260 -- 500~GeV, so it is
  complementary to the 4$b$ channel, more efficient for Higgs
  bosons having masses greater than 500~GeV.
  We analyse this di-leptonic decay channel in detail using
  kinematic variables such as $M_{\rm T2}$ and the $M_{\rm
    T2}$-assisted on-shell reconstruction of invisible momenta. Using
  proper cuts, a significance of $\sim$ 3$\sigma$ for 3000~fb$^{-1}$
  can be achieved at the 14~TeV LHC for $m_H = 260$ -- 400
  GeV if the mixing is close to its present limit and $\BR(H \to hh)
  \approx 1$. Smaller mixing values would require combining various
 decay channels in order to reach a similar significance.
 The complementarity among $ H \to hh$, $ H \to  ZZ$ and $ H \to WW$
 channels is studied for arbitrary $\BR(H \to hh)$ values.
}


\begin{document}

\begin{flushright}
   FTUAM-15-2\\
  IFT-UAM/CSIC-15-004\\
  KIAS-P15003
\end{flushright}

\vskip-6.0em
\maketitle
\vspace{-0.48cm}

\section{Introduction}

The discovery at the LHC by ATLAS and CMS collaborations of a scalar
boson compatible with the Standard Model (SM) Higgs
boson~\cite{Aad:2012tfa,Chatrchyan:2012ufa} has opened a new era in
particle physics.
Since there are several Higgs production modes and five of its decay
processes have been measured ($\gamma \gamma$, $ZZ^\ast$, $WW^\ast$,
$\tau \bar{\tau}$, and $b {\bar b}$), it is possible to extract its
couplings and compare with SM predictions. A useful variable to
evaluate the consistency of the experimental data with the SM Higgs
hypothesis is the so-called signal-strength modifier defined as
${\hat \mu} =\sigma_{\rm observed} /\sigma_{\rm SM}$ for each channel.
In the latest analyses of the full 7 and 8 TeV LHC data,
ATLAS~\cite{Aad:2014aba} and CMS~\cite{Khachatryan:2014jba} obtained following
combined values
\begin{align}
  \begin{array}{rcll}
    {\hat \mu_{\text{ATLAS}}} & = & 1.30^{+0.18}_{-0.17}\; & (m_h = 125.36~\text{GeV}),\\
    {\hat \mu_{\text{CMS}}}  & = & 1.00 \pm  0.13  & ( m_h = 125.02~\text{GeV})
  \end{array}
\end{align}
from the main Higgs production and decay modes. If we ignore the small
difference in the $m_h$ value used in the two fits\footnote{Both $m_h$
  and $\hat \mu$ should be fitted simultaneously.} and assume that
there are no correlation and gaussian error, the combined value is given
by\footnote{For a recent analysis including Tevatron data and taking
  into account the correlation among the different measurements,
  see~\cite{Bechtle:2014ewa}.
 } 
\begin{align}
  {\hat \mu_{\text{ATLAS+CMS}}} = 1.10 \pm 0.10 .
  \label{eq:StrengthModifier}
\end{align}
Thus, the experimental data is certainly consistent with the SM
predictions and it can be used to constrain new physics.

The simplest modification to the Higgs couplings is given by a generic
rescaling of them and this would be the case in the minimal extension
of the Higgs sector, in which a scalar singlet that generically mixes
with the ordinary SM Higgs is included. This is the model that we will
analyze in this work.
Adding a singlet to the SM scalar sector has implications that have
been widely explored in the literature. It can help to stabilize the
Higgs potential at high energies through their positive contributions
to renormalization group equations that govern the Higgs quartic
coupling evolution~\cite{EliasMiro:2012ay}. It can rescue the
electroweak baryogenesis scenario by providing a strong enough
first-order electroweak phase transition, as studied in
refs.~\cite{Espinosa:2011ax,Cline:2012hg} (see however,
ref.~\cite{Damgaard:2013kva}). Moreover, it can act as a dark matter (DM)
candidate~\cite{McDonald:1993ex} or as a portal to
DM~\cite{Baek:2012uj,Kim:2008pp,LopezHonorez:2012kv,Fairbairn:2013uta},
depending on its stability.

If the new scalar is not too heavy, it can be produced at the LHC
through the mixing with the ordinary Higgs and detected by the
conventional heavy SM-like Higgs boson
searches~\cite{TheATLAScollaboration:2013zha,Chatrchyan:2013yoa,ATLAS:2013nma,CMS:2013pea,CMS:xwa,CMS:bxa}.
On the other hand, if the double-Higgs decay mode is open, it will
decrease the significance of SM-like Higgs signatures. Consequently,
it is important to explore the specific resonant double-Higgs production \cite{Bowen:2007ia, Dolan:2012ac, No:2013wsa, Pruna:2013bma, Chen:2014ask}. 
In this work we extend previous analysis, focusing on the particular $hh\to b \bar{b} WW \to b \bar{b} \ell^+ \nu \ell^- \bar{\nu}$ process,  and present strategies to enhance the signal-background ratio by using various kinematic variables.

The organization of our paper is as follows. In
Section~\ref{sec:model} we briefly describe the model. Then, we review
the present constraints on the mass of the new singlet and its mixing
in Section~\ref{sec:constraints}. In turn, we study the production of
the heavy scalar and explore its detection in the next LHC run using the
double-Higgs decay channel with $b \bar{b} \ell^+ \nu \ell^-
\bar{\nu}$ as a final state. We comment on
the complementarity of this channel and the decays into two electroweak
gauge bosons. The interplay between direct production and indirect
effects, such as the modification of the Higgs couplings, will be
considered for a luminosity of 3000~fb$^{-1}$ in Section~\ref{sec:collider}.
In the next section we check the validity of our
study when extending the model to accommodate for a DM
candidate. Finally, we present the conclusions.

\section{The singlet-extended model}
\label{sec:model}

One of the simplest extension of the SM Higgs sector is given by the
addition of a real singlet field. This model has been also widely studied in
refs.~\cite{Hill:1987ea,Veltman:1989vw,Binoth:1996au,Schabinger:2005ei,
Patt:2006fw,Bhattacharyya:2007pb,Bowen:2007ia,Barger:2007im,
Barger:2008jx,Dawson:2009yx,Bock:2010nz,Baek:2011aa,
Fox:2011qc,Englert:2011yb,Englert:2011us,Batell:2011pz,Englert:2011aa,
Gupta:2011gd,Bertolini:2012gu,Dolan:2012ac,Batell:2012mj,Pruna:2013bma}.
The relevant Lagrangian for the scalar sector is as follows:
\begin{align}
  {\cal L} = (D_\mu \Phi)^\dagger (D^\mu\Phi) +
  \frac{1}{2} \partial_\mu S \partial^\mu S - V(\Phi,\, S),
\end{align}
with the potential~\cite{Schabinger:2005ei,Bowen:2007ia}
\begin{align}
  V(\Phi,\, S) =&~ \lambda_{40} \left( \Phi^\dagger \Phi - v^2
  \right)^2 + \lambda_{21} \, v \left( \Phi^\dagger \Phi - v^2 \right)
  S + \lambda_{22} \left( \Phi^\dagger \Phi - v^2 \right) S^2
  \nonumber\\
  & + \lambda_{02} \,v^2 \,  S^2 + \lambda_{03} \, v S^3 +
  \lambda_{04} S^4 .
  \label{eq:l1}
\end{align}
The physical doublet and singlet scalar fields can be obtained by
expanding the scalar potential $V(\Phi, \, S)$ around the real neutral
vacuum expectation values (VEVs):
\begin{align}
  \Phi = \left ( \begin{array}{c}
      0 \\ v + \phi/\sqrt{2}
  \end{array}
  \right), \quad S = v_S + s .
\end{align}
We take $v \simeq 174$~GeV and have chosen $V(\Phi,\, S)$ such
that $v_S = 0$.\footnote{Note that, in a generic potential, $S$ can be
  shifted to fulfill this condition.}
The conditions $\lambda_{40} >0$, $\lambda_{04} > 0$, and
$\lambda_{22} > - 2 \sqrt{ \lambda_{40} \lambda_{04}}$ have to be
imposed in order to ensure that the potential is bounded from below.

Due to the $\lambda_{21}$ term, the two scalars $\phi$ and $s$ mix and
the mass eigenstates are given by
\begin{align}
  \left (
  \begin{array}{cc}
    h \\ H
  \end{array}
  \right) = \left(
  \begin{array}{rr}
    \cos\alpha & \sin\alpha\\
    - \sin\alpha & \cos\alpha
  \end{array} \right) \left(
  \begin{array}{cc}
    \phi \\ s
  \end{array}
  \right) .
\end{align}
The mixing angle $\alpha$ and the mass eigenvalues read
\begin{align}
\begin{array}{rcl}
 \tan{2 \alpha} & = & \displaystyle \frac{\lambda_{21} \sqrt{2}} {2
   \lambda_{02} -\lambda_{40}} \\[3mm]
 m^2_{H,h} &  = & \left ( 2 \lambda_{40} + \lambda_{02} \pm \sqrt{2
     \lambda_{21}^2  + (2 \lambda_{40}  - \lambda_{02})^2 } \right)
 v^2 .
\label{eq:PhysicalParameters}
\end{array}
\end{align}
The stability of the vacuum requires $\lambda_{02} > 0$ and $4
\lambda_{40} \lambda_{02} > \lambda_{21}^2$.
We can use (\ref{eq:PhysicalParameters}) to express
$(\lambda_{40},\, \lambda_{02}, \, \lambda_{21})$ in terms of the
physical parameters $\alpha$, $m_h$, $m_H$, and $v$ as follows:
\begin{align}
  \begin{array}{rcl}
    \lambda_{40}   & = &   \displaystyle \frac {m_h ^2 \sin^2{\alpha}   + m_H ^2 \cos^2{\alpha} }{8v^2},  \\[3mm]
    \lambda_{02}   & = &   \displaystyle \frac {m_h ^2 \cos^2{\alpha}   + m_H ^2 \sin^2{\alpha} }{4v^2},  \\[3mm]
    \lambda_{21}   &  = &     \displaystyle \frac {(m_H ^2-  m_h ^2) \sin{2\alpha} }{2 \sqrt{2}v^2}  .
  \end{array}
\end{align}
The cubic and quartic interactions involving the mass eigenstates $h$,
$H$
can be given as functions of the physical
parameters~(\ref{eq:PhysicalParameters}) and the three remaining
couplings $(\lambda_{22}, \,\lambda_{03}, \,\lambda_{04})$. This is in
contrast with the SM (or in the extended complex Higgs singlet
model), where the full potential can be reconstructed from the mass
(matrix) and the VEVs of the field(s).
In what follows, we assume that $h$, the lighter Higgs, is the
SM-like Higgs discovered at the LHC having $m_h \sim 125$~GeV. Its
couplings approach the SM ones in the $\cos\alpha \approx 1$
limit.

\section{\boldmath Constraints on $m_H$ and $\sin\alpha$}
\label{sec:constraints}

\begin{figure}[ht]
  \begin{center}
    \includegraphics[scale=0.75]{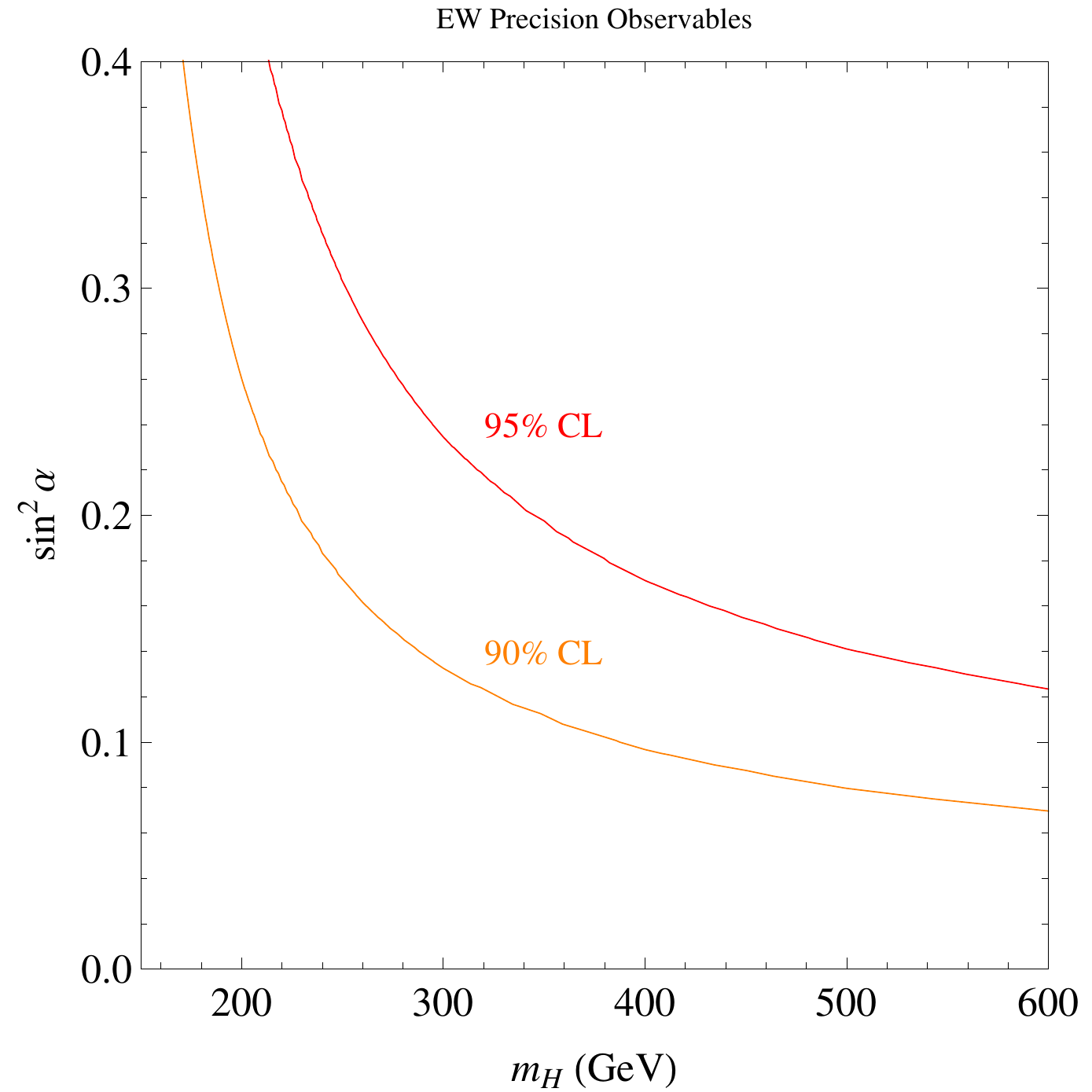}
  \end{center}
  \caption{Constraints in the  $m_{H} - \sin^2\alpha$ derived from the
    full set of EWPO at the $Z$-peak.
    }
\label{fig:EWPT}
\end{figure}

The deviation of the Higgs couplings from their SM values is
constrained by the LHC Higgs data and by the electroweak precision
observables (EWPO). We first concentrate on the latter.
The Higgs contributes to the gauge bosons self-energies involved in
the EWPO.
In the singlet-extended Higgs model, the one loop self-energies will be given
by the sum of two SM-like Higgs contributions evaluated at Higgs
masses, $m_h$ and $m_H$,  rescaled by $\cos^2 \alpha$ and $\sin^2
\alpha$ respectively~\cite{Bowen:2007ia}.
This property can also be applied to non-universal diagrams (\eg,
vertex corrections) involving the Higgses and it is transmitted  to
the EWPO in the limit where higher order, ${\cal O} (\sin^4 \alpha)$
terms, are neglected. Taking this into account and using
ZFITTER~\cite{Bardin:1989di,Bardin:1989tq,Bardin:1990de,Bardin:1992jc,
  Bardin:1999yd,Arbuzov:1999uq,Arbuzov:2005ma,ZFITTER:6.43},
we have evaluated the predictions for the $Z$-peak
observables~\cite{ALEPH:2005ab} and $m_W$,
$\Gamma_W$~\cite{Schael:2013ita}, as a function of $m_H$ and $\sin^2
\alpha$. The list of observables used are listed in Table~\ref{tab:ewpo}.
\begin{table}[!h]
  \begin{center}
    \scalebox{0.85}{
      \begin{tabular}{| c c c c |}
        \hline
        Observable  & Data & Observable & Data \\
        \hline
        \hline
        $m_W$&$80.385\pm 0.015$&$\sin^2\theta_{\rm eff}^\ell$&$0.2324\pm 0.0012$\\
        $\Gamma_W$&$2.085\pm 0.042$&$A_c$&$0.670\pm 0.027$\\
        $\Gamma_Z$&$2.4952\pm 0.0023$&$A_b$&$0.923\pm 0.020$\\
        $\sigma^0_{\rm had}$&$41.540\pm 0.037$&$A_{\rm FB}^{0,c}$&$0.0707\pm 0.0035$\\
        $R^0_\ell$&$20.767\pm 0.025$&$A_{\rm FB}^{0,b}$&$0.0992\pm 0.0016$\\
        $A_{\rm FB}^{0,\ell}$&$0.0171\pm 0.0010$&$R^0_c$&$0.1721\pm 0.0030$\\
        $A_{\ell}$&$0.1499\pm 0.0018$&$R_b^0$&$0.21629\pm 0.00066$\\
        \hline
      \end{tabular}
    }
  \end{center}
  \caption{Electroweak data taken from ref.~\cite{Baak:2014ora} used
    in the fit of the EWPO.}
\label{tab:ewpo}
\end{table}
The results are presented in Figure~\ref{fig:EWPT}, where 90\% and
95\% C.L. allowed regions in the $m_H -\sin^2 \alpha$ plane are
shown. The structure of these lines can be understood by noting that
at $m_H = m_h$ the contour line is a vertical one since its value does
not depend on the mixing angle. On the other hand, for large $m_H$
values, the mixing angle must be small enough to compensate the
non-decoupling Higgs contributions to the EWPO.

It is also common to use the oblique parameters $(S, \, T, \, U)$
instead of analyzing the complete set of observables. We expect that
in the region where $m_H \lesssim 200$~GeV both methods should give a
similar $\chi^2$ value. However, for larger $m_H$ values, the gaussian
approximation to the $\chi^2$ that is used to fit  $(S, \, T,\, U)$ and
the estimation of their errors starts to break down.\footnote{This is
  shown in~\cite{Lopez-Val:2014jva}, where a detailed  calculation of
  $\Delta r$ and $m_W$ in the singlet-extended model is presented.}
This can be explicitly checked by evaluating $\chi^2$ as a function of
$m_h$ using the whole $Z$-peak data or the oblique parameters $(S,\,
T,\,U)$.

\begin{figure}[ht]
  \begin{center}
    \includegraphics[scale=0.7]{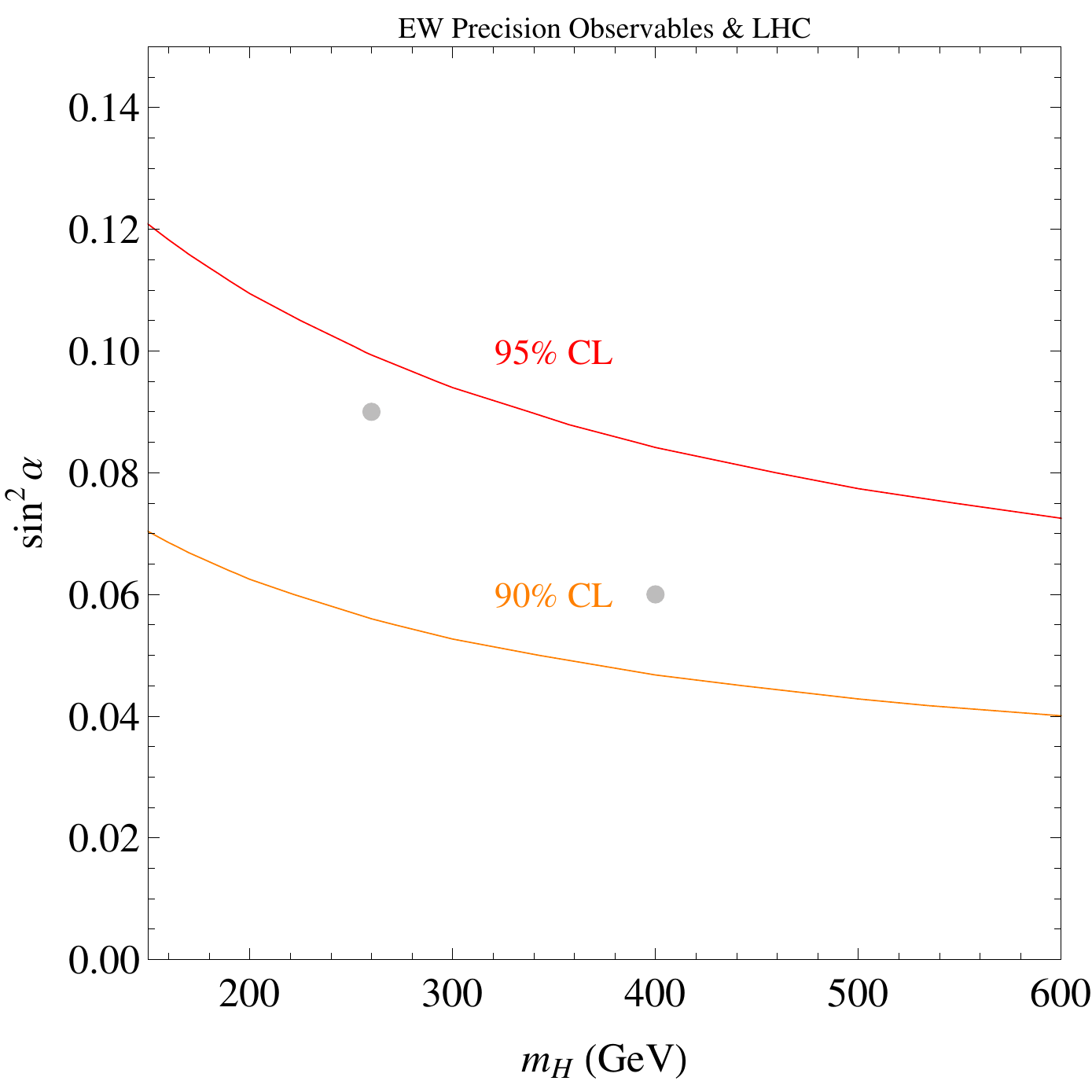}
   \end{center}
  \caption{Constraints in the $m_{H}-\sin^2\alpha$ plane
    derived from the full set of EWPO  at the $Z$-peak combined with the LHC Higgs
    coupling data. We have also drawn the two benchmark points
    whose LHC implications are analyzed in detail in
    Section~\ref{sec:collider}.}
\label{fig:LHCBounds}
\end{figure}

Let us now consider the impact of the LHC Higgs data. As already
mentioned in the introduction, the reduction of the Higgs couplings to
SM fields due to the mixing  translates into a common reduction of the
Higgs signal-strength modifier in all channels. By applying the
narrow-width approximation, one can see that this factor is given by
$\cos^2{\alpha}$.
Using eq.~(\ref{eq:StrengthModifier}), it is straightforward to derive
the following bound on the mixing:
$\sin^2 \alpha < 0.076$ (0.128) at 90\% (95\%) C.L.
We can now combine this bound with  the ones derived from the EWPO: the results are given in
Figure~$\ref{fig:LHCBounds}$.

\begin{figure}[ht]
  \begin{center}
    \includegraphics[scale=0.7]{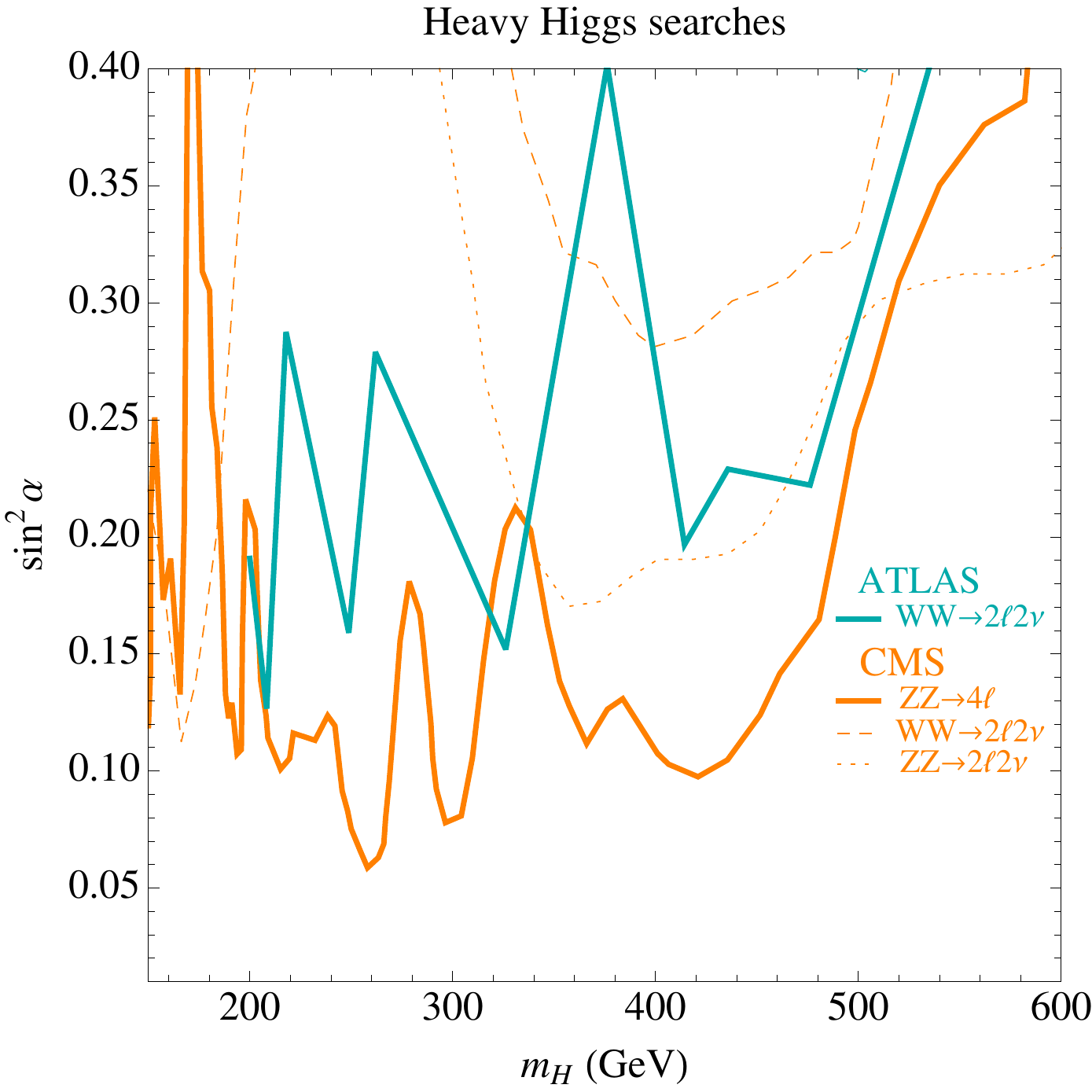}
  \end{center}
  \caption{
    Constraints in the  $m_{H} -\sin^2\alpha$ plane derived by ATLAS and
    CMS from SM-like heavy Higgs searches assuming the heavy
    Higgs decays as the SM one. For non-zero \BR$(H \to hh) $ values, the
    vertical axis would read  $\sin^2 \alpha / (1 -  \BR (H \to hh))$.
  }
\label{fig:CMSBounds}
\end{figure}

After dealing with the indirect bounds\footnote{There are other
  constraints that can be derived by imposing perturbative unitarity
  of scattering amplitudes for longitudinal $W$
  bosons~\cite{Lee:1977eg,Englert:2011yb}. We will ignore them since
  they are weaker than the other bounds~\cite{Baek:2012uj}.}
on the mixing for a given $m_H$ value, we briefly comment on
the direct ones, derived from heavy Higgs boson searches.
 Note that, as a consequence of the mixing,
 the production and decay modes of the singlet-like Higgs $H$ will
 be the same as those of the SM-like Higgs.
 However, as it has different mass, the branching ratios of the
decay channels will be different from the SM-like Higgs. We
can reinterpret ATLAS and CMS analyses for heavy Higgs searches to
derive bounds on $m_H$ and $\sin^2\alpha$.
The ATLAS collaboration presented two searches for the heavy Higgs
boson. The first one uses $H \to WW \to \ell \nu \ell
\nu$~\cite{TheATLAScollaboration:2013zha} decay mode and the bound
corresponds to an integrated luminosity of 21~fb$^{-1}$ at $\sqrt{s}
= 8$~TeV. The second one uses the $H \to ZZ$
decay~\cite{ATLAS:2013nma}.
The CMS collaboration has reported two analyses on heavy Higgs
searches using the $H \to  ZZ$ decay channels. The first one
corresponds to integrated luminosity 19.6~fb$^{-1}$ at
$\sqrt{s} = 8$ TeV and considers the $\ell^+\ell^- q \bar q$ final
state~\cite{CMS:2013pea}. The second one considers final states where
both $Z$'s decay into charged leptons and corresponds to integrated
luminosity 5.1~fb$^{-1}$ at $\sqrt{s} = 7$ TeV and 19.6~fb$^{-1}$ at
$\sqrt{s} = 8$ TeV~\cite{CMS:xwa}.
The CMS collaboration has also performed an analysis using the channel
$H \to WW \to \ell \nu \ell \nu$, obtained for the configurations of
$\sqrt{s}=7$ TeV with integrated luminosity of 4.9~fb$^{-1}$ and
$\sqrt{s}=8$ TeV with 19.5~fb$^{-1}$~\cite{CMS:bxa}.
The results are shown in Figure~\ref{fig:CMSBounds}, where we have
assumed that $H$ has the same branching ratios as a SM Higgs would have for those masses.
 This is certainly a good approximation if the $H \to hh$
decay process is not kinematically allowed,
or $\BR(H \to hh) \ll 1$. On the other hand, if $\BR(H
\to hh)$ is substantially large, these bounds have to
be rescaled as indicated in the figure, and eventually will become
 irrelevant in the $\BR(H \to hh) \sim 1 $ limit.
In this case, the double-Higgs production process will be the main signature of the
model at the LHC and deserves a detailed study. We investigate the
scenario in the next section, and in turn, present the interplay among
the different, present and future, bounds.

\section{\boldmath Detecting the heavy Higgs in $H \to hh$ at the LHC}
\label{sec:collider}

\begin{figure}[ht]
  \begin{center}
    \includegraphics[scale=0.53]{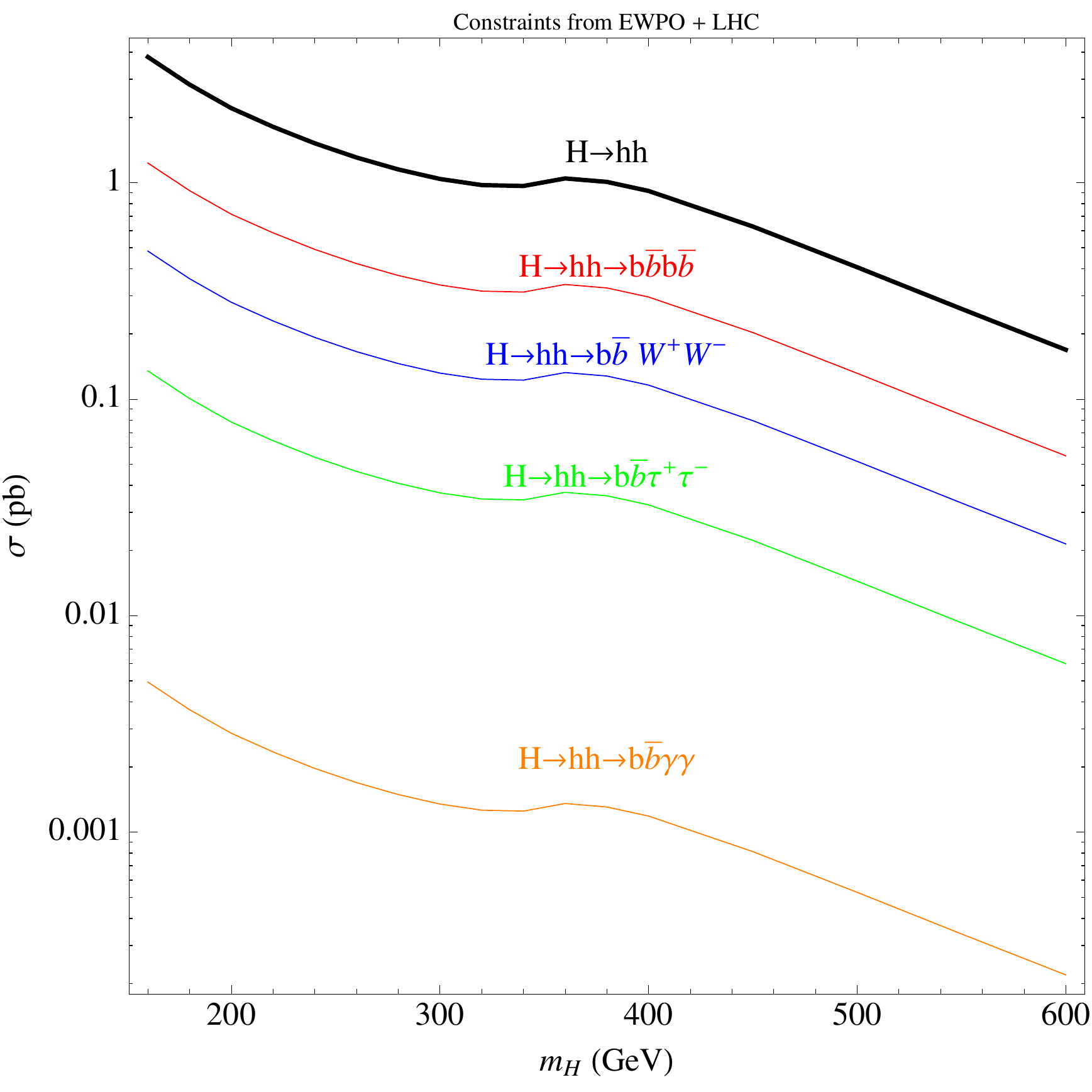}
  \end{center}
  \caption{Cross-section of the double Higgs production mediated by
    the heavy Higgs boson evaluated at the maximal mixing angle.
    The black line shows the total cross
    section for this process while different final state cross
    sections are presented in colours.
      }
\label{fig:xsH}
\end{figure}

\noindent
The resonant double Higgs production is a distinct feature of the
model we are dealing with. In this section we study this process in
the forthcoming LHC run at 14 TeV. Since the Higgs production
cross-section $\sigma (H)$ scales as $\sin^2 \alpha$  and there is a
bound on the allowed mixing for a given $m_H$, we can obtain the
maximal value of $\sigma (H)$ as a function $m_H$. This is shown
in Figure~\ref{fig:xsH}, where the $95 \%$ C.L. limit on $\sin^2
\alpha$ has been used.

In order to check the signal significance at the LHC, which will be resumed with the upgraded center-of-mass
energy  along the year 2015, we perform a Monte Carlo (MC) study by
choosing two benchmark points.
For the $H \to hh$ decay process, the largest
portion of signal events will consist of the four-$b$-jet final
state as studied in
refs.~\cite{Cooper:2013kia,deLima:2014dta}. However, the multi-jet
signature is generically vulnerable to the
huge QCD backgrounds and the poor reconstruction efficiency. One can
attempt to increase the purity of signal events by imposing a tight
$b$-tagging condition, but then it would result in a big
sacrifice of the signal statistics.
The ATLAS and CMS collaborations have performed  searches for resonant
double-Higgs production in the $b\bar{b}b\bar{b}$ final
state~\cite{AtlasConf2014005, CMS:2014eda}. It was shown that in order to be
effective in this channel the mass of the new resonance should be
greater than 500 GeV to ensure two highly boosted, back-to-back
$b\bar{b}$ di-jet systems~\cite{Cooper:2013kia}.
For smaller masses, the acceptance times the efficiency of the search
decreases, thus making difficult to use this channel.

The subleading decay process is $b\bar{b} W^+W^-$, followed by
fully-hadronic, semi-leptonic, and di-leptonic modes. This search
channel, as it will be shown below, can be efficient in the range of the
heavy Higgs mass 260 -- 500 GeV.\footnote{The complementary channel to the
  one presented here is the $b\bar{b}\gamma \gamma$~\cite{Aad:2014yja, CMS:2014ipa}. The small
  branching ratio of the SM Higgs decaying into two photons makes this
  channel challenging (see,
  however refs.~\cite{Barger:2014taa,Barger:2014qva}).} Among them, the
final states containing the lepton are more suitable for the
search since the fully-hadronic states are liable to be in trouble due
to the similar reason as in the four-$b$-jet signal.
In leptonic signal events the missing energy
originated from the neutrino prevents the direct reconstruction
of the Higgs resonances.
Still, provided that the light Higgs boson mass $m_h$ is
accurately known, one can obtain the neutrino four-momenta up to a
two-fold ambiguity by using on-shell mass relations as well as the
missing energy condition in the case of the semi-leptonic channel:
\begin{align}
  \left ( p^\nu \right )^2 = 0, \quad
  \left ( p^\nu + p^\ell + p^q + p^{q^\prime} \right )^2 = m_h^2, \quad
  \mathbf{p}_{\rm T}^\nu = \slashed{\mathbf{p}}_{\rm T} ,
\end{align}
where $\slashed{\mathbf{p}}_{\rm T}$ is the missing transverse
momenta measured in the event, and $q$ and $q^\prime$ are jets from
the hadronically-decaying $W$ boson.
On the other hand, the on-shell relations are not enough to
constrain the neutrino momenta in the case of the di-leptonic
channel even though it provides cleaner signal than the
semi-leptonic one.
We here examine the discovery potential of the di-leptonic decay mode,
which appears to be more challengeable due to the missing neutrinos although
it is less vulnerable to uncertainties regarding the jet
reconstruction, by using various kinematic variables and an
approximate reconstruction scheme. We will see the practicability and
the limitation of the search strategy in two different scenarios
characterized by
\begin{enumerate}
  \item $m_H = 400$ GeV, $\sin^2\alpha = 0.06$,
  \item $m_H = 260$ GeV, $\sin^2\alpha = 0.09$,
\end{enumerate}
and assume $\BR (H \to hh) = 1$ for both benchmark
points.\footnote{We stress that the results obtained here can be
  readily reinterpreted for the scenario with different $\BR (H \to
  hh)$ values.}

The production cross-section is $\sigma (gg \to H) \times \BR(H \to
hh) = \sigma(gg \to \phi) \times \sin^2\alpha \simeq 0.7$ (1.2) pb for
$m_H = 400$ (260) GeV in the 14-TeV LHC condition.
Here, $\phi$ is the Higgs-like scalar and  $\sigma(gg \to \phi)$
values have been obtained from the Higgs Cross Section Working Group
in~\cite{Dittmaier:2011ti} assuming that the couplings of $\phi$ are
SM-like. The search channel of interest is
\begin{align}
  H \to hh \to b \bar{b} \, W^+W^- \to b (p^b) \bar{b} (p^{\bar{b}}) +
  \ell^+ (p^{\ell}) \ell^- (q^{\ell}) +
  \slashed{E}_{\rm T} \quad (\ell = e,\,\mu),
  \label{eq:signal_topology}
\end{align}
where the source of the missing energy is the neutrinos produced by
the leptonically-decaying $W$ bosons. For a numerical analysis,
we have generated the MC events using
\textsc{Pythia 8}~\cite{Sjostrand:2006za}, interfacing with the CT10
parton distribution functions~\cite{Lai:2010vv} for a proton-proton
collision at $\sqrt{s} = 14$~TeV. The parton showering
and the hadronization have been performed by \textsc{Pythia 8}. Then,
the hadron-level data has been processed with the fast
detector-simulation program \textsc{Delphes
  3}~\cite{deFavereau:2013fsa}, which reconstructs the final-state
objects such as jets, isolated leptons, and the missing energy with the
inclusion of detector resolution effects and tagging/fake rates. The
input parameters have been adjusted for the ATLAS detector
in \textsc{Delphes}. \textsc{FastJet 3}~\cite{Cacciari:2011ma} is
employed to reconstruct jets. In our simulation, the anti-$k_t$ jet
algorithm~\cite{Cacciari:2008gp} with distance parameter of
0.5 is chosen for the jet reconstruction.
It is known that the tagging efficiency for the $b$-jet depends on the
transverse momentum $p_{\rm T}$ and the pseudorapidity $\eta$ of
the jet object. Recent ATLAS and CMS analyses on the $b$-jet
identification for the experimental data indicate that the
efficiency can be as large as $\sim 60$ --
80\%~\cite{ATLAS:2012lma}. For the sake of a simple analysis,
we assume a flat $b$-tagging efficiency of 70\% for $p_{\rm T} > 30$
GeV and $|\eta| < 2.5$ and the mis-tagging efficiency is set to be 10\% for
the $c$-jet and 1\% for the light flavor as well as the gluon jet.
Isolated electrons (muons) are required to have $p_{\rm T} > 13$ (10)
GeV within $|\eta| < 2.4$. In order to remove fake leptons coming from
the decays of hadrons, we discard the leptons lying within the angular
separation $\Delta R_{\ell j} = \sqrt{\Delta \phi_{\ell j}^2 +
  \Delta \eta_{\ell j}^2} < 0.4$ from a jet with $p_{\rm T} > 25$
GeV. Since the tau reconstruction efficiency is relatively
poor, we reject the events containing the tau-jet with $p_{\rm T} >
10$ GeV.
The missing transverse momentum $\slashed{\mathbf{p}}_{\rm T}$ is
defined as the opposite of the vector sum of all the visible
particles' transverse momenta.

Having the same final states that the signal, the
di-leptonic $t\bar{t}$ process is the main background. The subleading
backgrounds include Drell-Yan (DY), di-boson, and the SM
Higgs processes that lead into the leptonic final states and the
$b$-jets.
In addition, we consider the
rare SM Higgs processes including the double-Higgs production via the
gluon-gluon fusion (GGF), the single-Higgs production via the
vector-boson fusion (VBF), and the Higgs boson produced in
association with a weak vector boson or a top-pair, \ie, $h W/Z$ and $h
t\bar{t}$. The SM double-Higgs events have been generated by a modified
\textsc{Pythia 6} program~\cite{El-Kacimi:2002qba}  
in which  the matrix element calculated with \textsc{hpair}~\cite{Dawson:1998py}
is implemented, while the other processes have
been generated by \textsc{Pythia 8}. We use the production cross
section for the SM double-Higgs process obtained by \textsc{hpair},
which can calculate up to a next-to-leading order value. The $t\bar{t}$
production cross section is calculated with
\textsc{Top++ 1.4}~\cite{Czakon:2011xx} at next-to-next-leading order,
and the Higgs production cross sections except that of the
double-Higgs process are obtained from ref.~\cite{Dittmaier:2011ti}.
For the DY and the di-boson
processes, we use the leading-order cross sections calculated with
\textsc{Pythia 8} since most of them can contribute to the background
by faking $b$-jets and can be readily removed by event selection cuts,
which will be discussed shortly. In Table~\ref{tab:cross_section}, we
show the cross-section values used in this study.
\begin{table}[bt!]
  \begin{center}
    \begin{tabular}{c | c}
      \hline&\\[-2mm]
      Process & Cross section\\[2mm]
      \hline&\\[-2mm]
      $H \to hh$ ($m_H = 400$ GeV) & 0.66\\[2mm]
      $H \to hh$ ($m_H = 260$ GeV) & 1.18\\[2mm]
      \hline&\\[-2mm]
      $t\bar{t}$ & 844.43\\[2mm]
      GGF $h$ & 50.35\\[2mm]
      VBF $h$ & 4.17\\[2mm]
      $hW/Z$ & 2.39\\[2mm]
      $h t\bar{t}$ & 0.61\\[2mm]
      $hh$ & 0.033\\[2mm]
      DY & 91130.0\\[2mm]
      Di-boson & 121.0\\[2mm]
      \hline
    \end{tabular}
  \end{center}
  \caption{
    Production cross sections in pb for the signal and background
    processes at the 14 TeV proton-proton collision. We set $m_t =
    173.5$~GeV and $m_h = 125$~GeV.
  }
\label{tab:cross_section}
\end{table}

Before going further into the analysis, we introduce one of the main
kinematic variables and the
reconstruction scheme adopted to obtain the approximate values of
the invisible neutrino momenta. The situation with more than one
invisible particle in a collider event is common in many extensions
of the SM providing a viable DM candidate. One of the
most studied collider variables to search for such a new physics
signature is $M_{\rm T2}$, which is a generalized transverse mass
particularly known to be useful for the pair-production processes of
new particles that eventually decay into the invisible
particles~\cite{Lester:1999tx, Barr:2003rg}.
Suppose that the decay topology is like
\begin{align}
  p p \to Y + \overline{Y} + U \to V (p) \chi (k) + \overline{V} (q)
  \chi (l) + U (u),
  \label{eq:pair_decay_topology}
\end{align}
where $Y$ is a heavy unstable particle, $V$ is a set of
detectable particles such as jets or charged leptons, and $\chi$ is the
invisible particle. Here, $U$ denotes a set of particles that does not
participate in the decay process of $Y$ like initial or final state
radiations.
For the new physics signature with the decay
topology~(\ref{eq:pair_decay_topology}), the invisible
momenta $k$ and $l$ as well as the particle masses $m_Y$ and $m_\chi$
are unknown, while only the sum of their
transverse components can be inferred from the deficit of total
transverse momentum in the collider event, \ie, the missing transverse
momentum. Then, $M_{\rm T2}$ is defined as
\begin{align}
  M_{\rm T2} \equiv \min_{\mathbf{k}_{\rm T} + \mathbf{l}_{\rm T} =
    \slashed{\mathbf{p}}_{\rm T}}
  \left [
    \max \left \{
      M_{\rm T}^{(1)}, \, M_{\rm T}^{(2)}
      \right \}
  \right ],
  \label{eq:m_T2_def}
\end{align}
where $M_{\rm T}^{(i)}$ ($i=1$, 2) are transverse masses for the decay
chains,
\begin{align}
  \left ( M_{\rm T}^{(1)} \right )^2 & =
  m_V^2 + m_\chi^2 + 2\left( \sqrt{|\mathbf{p}_{\rm T}|^2 +
      m_V^2}
    \sqrt{|\mathbf{k}_{\rm T}|^2 + m_\chi^2} - \mathbf{p}_{\rm T} \cdot
    \mathbf{k}_{\rm T} \right ) , \nonumber\\
  \left ( M_{\rm T}^{(2)} \right )^2 & =
  m_{\overline{V}}^2 + m_\chi^2 + 2\left(
    \sqrt{|\mathbf{q}_{\rm T}|^2 +
      m_{\overline{V}}^2}
    \sqrt{|\mathbf{l}_{\rm T}|^2 + m_\chi^2} - \mathbf{q}_{\rm T} \cdot
    \mathbf{l}_{\rm T} \right ).
  \label{eq:m_T}
\end{align}
Here, $\mathbf{k}_{\rm T}$, $\mathbf{l}_{\rm T}$, and $m_\chi$ are
input parameters. In practice, the transverse momenta of
invisible particles are uniquely determined by the minimization,
whereas the invisible particle
mass $m_\chi$ is a constant that must be put by hand before the
minimization. Once the correct
$m_\chi$ value is chosen, the endpoint position of the $M_{\rm T2}$
distribution points to the parent particle mass,
\begin{align}
  M_{\rm T2} (m_\chi = m_\chi^{\rm true}) \leq m_{Y} .
\end{align}
The situation becomes simpler when the invisible particle mass is
already known as in the case of SM processes, where the neutrino is the
only particle escaping detection and can be safely assumed to be
massless for reconstruction purposes.\footnote{The application of $M_{\rm T2}$ to the SM di-leptonic $t\bar{t}$ process was firstly proposed in~\cite{Cho:2008cu}.
It is later employed and checked its efficiency in the real
experimental analyses measuring the top quark mass at both
Tevatron and the LHC~\cite{Aaltonen:2009rm,ATLAS:2012poa,Chatrchyan:2013boa}.}
Another notable feature of the $M_{\rm T2}$ variable is that it comes
in handy even when one or both parent particles are off-shell. This
has been studied to measure the SM Higgs boson mass in the
di-leptonic $WW^{(\ast)}$ channel~\cite{Choi:2009hn,Choi:2010dw}. In the case when
$m_h < 2m_W$, at least one of the $W$ bosons should be produced
off-shell. Then, the maximal value of $M_{\rm T2}$ is not
$m_W$, but $\sim m_h / 2$. This can be deduced by considering some special
kinematic configurations as derived in Appendix~\ref{sec:m_T2_Higgs}.

As mentioned above, the di-leptonic system cannot be solved by
on-shell mass relations even if Higgs boson masses are
known. However, there is an approximation scheme to solve  the
unknown neutrino momenta by help of $M_{\rm T2}$.
When the minimization has been finalized to obtain the $M_{\rm T2}$
value, a unique solution for the transverse momentum configuration is
picked up.
\begin{figure}[t!]
  \begin{center}
   \includegraphics[width=0.48\textwidth]{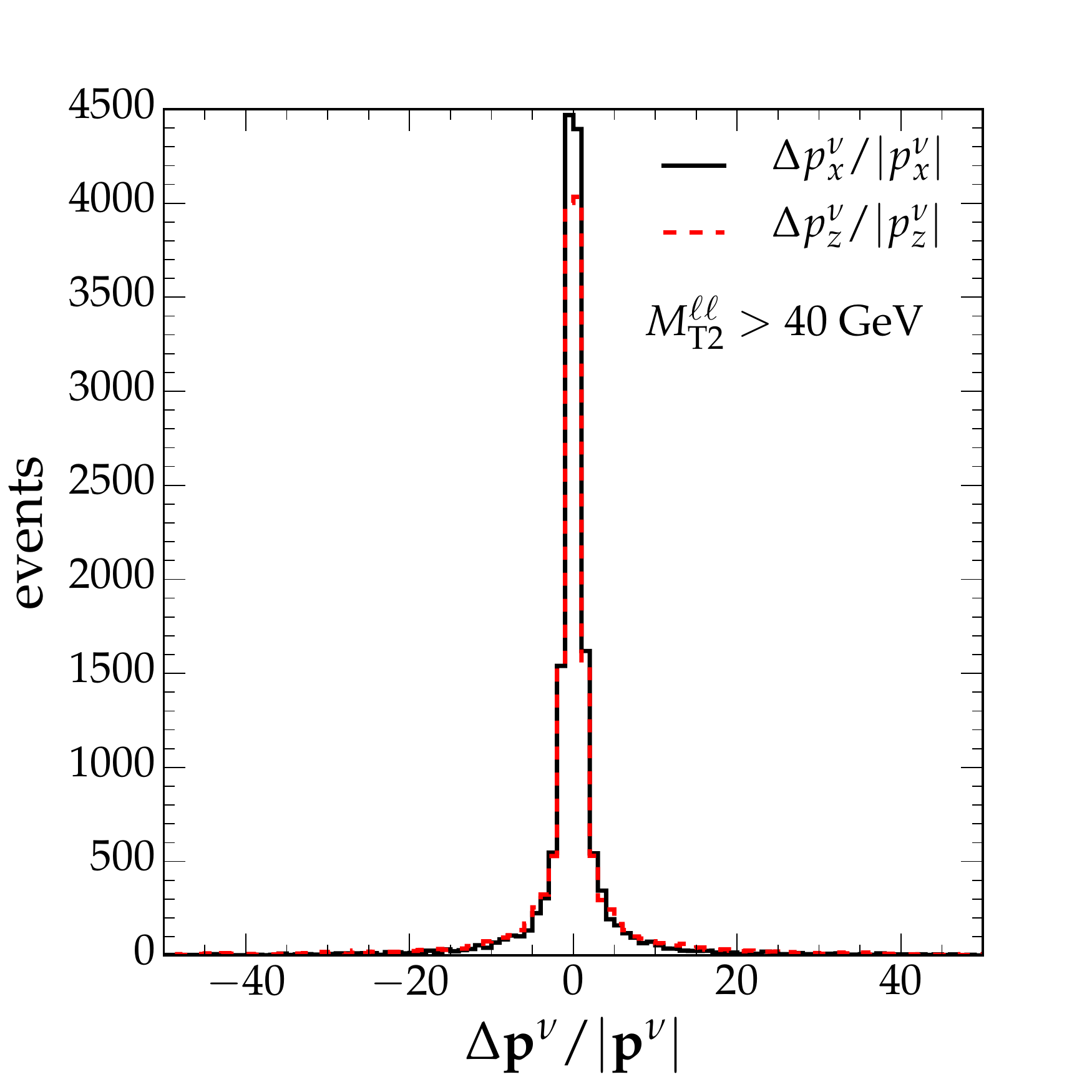}
   \includegraphics[width=0.48\textwidth]{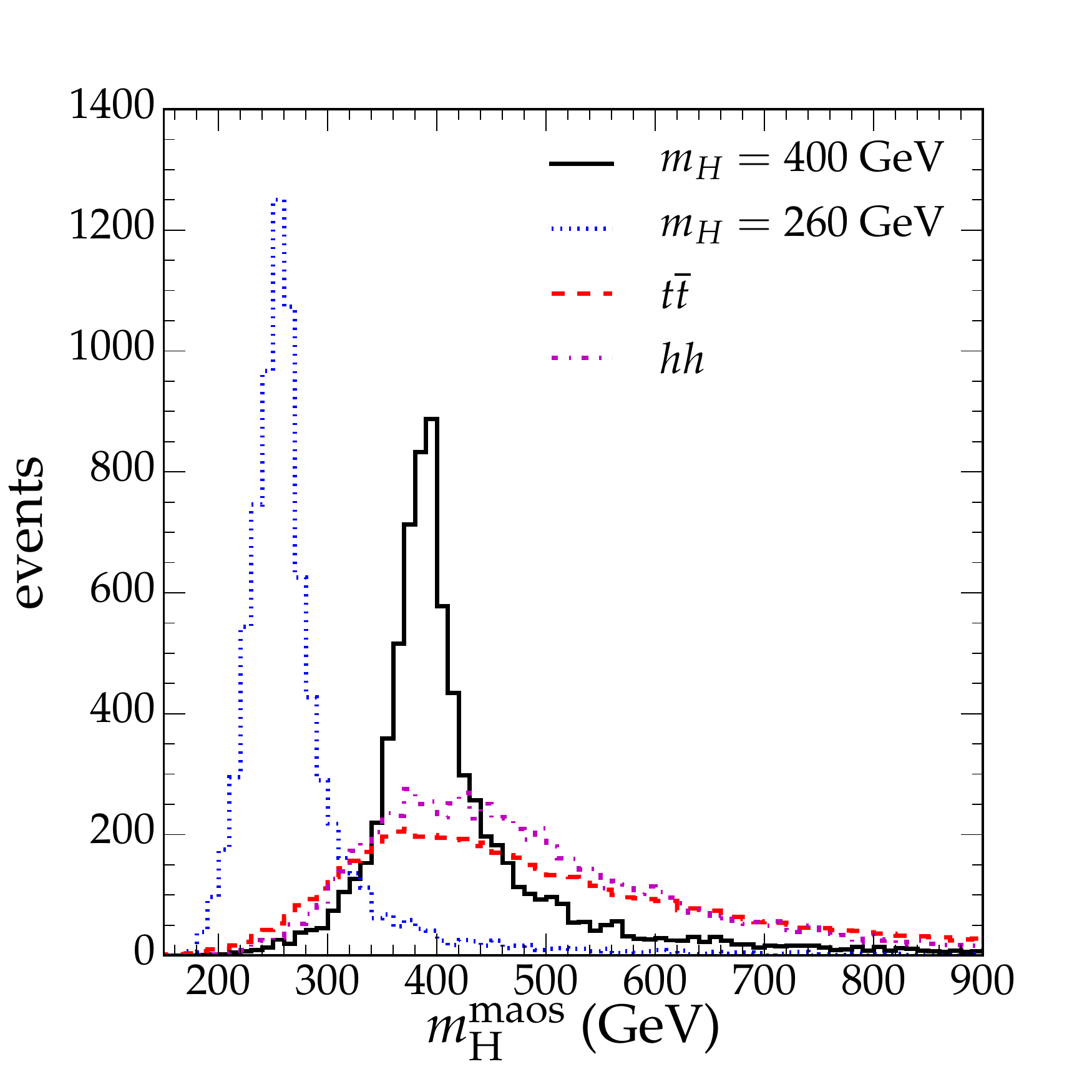}
      \end{center}
  \caption{(Left panel) Difference between the MAOS momentum and
    the true neutrino momentum for $m_H = 400$ GeV $\Delta
    \mathbf{p}^{\nu} / |\mathbf{p}^{\nu}| \equiv
    (\mathbf{p}_{\nu}^{\rm maos} - \mathbf{p}_\nu^{\rm
      true})/|\mathbf{p}_{\nu}^{\rm true}|$.
    (Right panel) Normalized $m_{H}^{\rm maos}$ distributions for $m_H
    = 260$ and 400 GeV and the $t\bar{t}$ backgrounds using
    parton-level data.}
\label{fig:maos_parton}
\end{figure}
One may attempt to see the correlation between these hypothetical
momentum components and the true ones.
For a subset of events whose $M_{\rm T2}$ values are close to $M_{\rm
  T2}^{\rm max}$, it can be shown that the $M_{\rm T2}$
solution of the transverse momenta are very close or equal to the true
momenta. This can be justified by the fact that the $M_{\rm T2}$
solution is unique while preserving kinematic constraints,\footnote{
  The $M_{\rm T}^{(i)}$ ($i = 1$, 2) functions are
  ellipses in the phase space and the $M_{\rm T2}$ value is
  determined by their intersecting point in the balanced
  configuration. This feature can be used to seek the $M_{\rm T2}$
  value by using a sophisticated algorithm proposed
  in~\cite{Cheng:2008hk}.
}
and the endpoint of the transverse mass corresponds to the
invariant mass of the decaying system, \ie, the parent particle mass.
Then, by adopting the $M_{\rm T2}$ solution of the invisible
transverse momenta in conjunction with known on-shell
mass relations, one can calculate the longitudinal and energy
components of the invisible four-momenta. This is so-called $M_{\rm
  T2}$-assisted on-shell (MAOS) approximation scheme~\cite{Cho:2008tj}.
One drawback of this scheme is that it cannot be applied if any
of the parent particles are off-shell.
 However, it was claimed that one can
circumvent the on-shell mass problem by substituting the transverse
mass for the decay chain instead of the invariant mass into the on-shell
mass relation~\cite{Choi:2009hn,Choi:2010dw,Park:2011uz}. It means
that the on-shell mass relations now become
\begin{align}
  \left( p + k^{\rm maos} \right)^2 = \left ( M_{\rm T}^{(1)} \right
  )^2, \quad
  \left( q + l^{\rm maos} \right)^2 = \left ( M_{\rm T}^{(2)} \right
  )^2.
  \label{eq:MAOS_modified}
\end{align}
This modified scheme guarantees that there is always a
real solution for the invisible momenta since the transverse mass is
bounded from above by the invariant mass, and it maintains the
property that the MAOS momentum is equal to the true momentum for the
endpoint events of $M_{\rm T2}$. See the left panel of
Figure~\ref{fig:maos_parton}, where the efficiency of
approximation to the invisible momenta in the modified scheme is
shown. Since the light Higgs boson mass here is set to be 125 GeV $<
2m_W$, one or both $W$ bosons produced by the Higgs boson are always
off-shell. In this situation, the modified MAOS scheme
(\ref{eq:MAOS_modified}) can be applied.
Once the MAOS momentum has been obtained, one can construct the
invariant mass of the total system, which corresponds to the heavy
singlet-like Higgs boson mass given by
\begin{align}
  \left(m_H^{\rm maos}\right)^2 \equiv \left( p^b + p^{\bar{b}} +
    p^{\ell} + q^{\ell} + k^{\rm maos} +
    l^{\rm maos} \right)^2 \simeq m_H^2 .
\end{align}
Strictly speaking, the equality holds only when $k^{\rm maos} = k^{\rm
  true}$ and $l^{\rm  maos} = l^{\rm true}$. The right panel of
Figure~\ref{fig:maos_parton}, which shows $m_{H}^{\rm maos}$ distributions
for the heavy Higgs signal and the SM double-Higgs as well as
$t\bar{t}$ backgrounds using the parton-level MC event samples. One
can see that the peak position of the signal distribution
clearly matches the $m_H$ value, while broad distributions are
exhibited in the non-resonant background process.

Armed with these tools, we now discuss our analysis to search for the
heavy Higgs signal. After reconstructing the final-state objects, we
select events that passed the basic cuts given as follows.
\begin{itemize}
  \item At least two isolated, opposite-sign leptons including the
    electron or the muon, \ie, $e^\pm e^\mp$,  $\mu^\pm \mu^\mp$, and
    $e^\pm \mu^\mp$. We further require that one of them must
    have $p_{\rm T} > 20$ GeV,
  \item At least two $b$-tagged jets with $p_{\rm T} > 30$ GeV,
  \item Missing energy $\slashed{E}_{\rm T} > 20$ GeV,
  \item For the opposite-sign same-flavor leptons, the event is
    rejected if $m_{\ell^+\ell^-} < 12$ GeV to avoid the leptons
    produced by decays of the hadrons, and the $Z$-veto condition, which
    discards events containing $|m_{\ell^+\ell^-} - m_Z| < 15$ GeV, is
    imposed.
\end{itemize}
We note that all the cut values have been chosen to optimize the
signal significance.
\begin{figure}[t!]
  \begin{center}
    \includegraphics[width=0.48\textwidth]{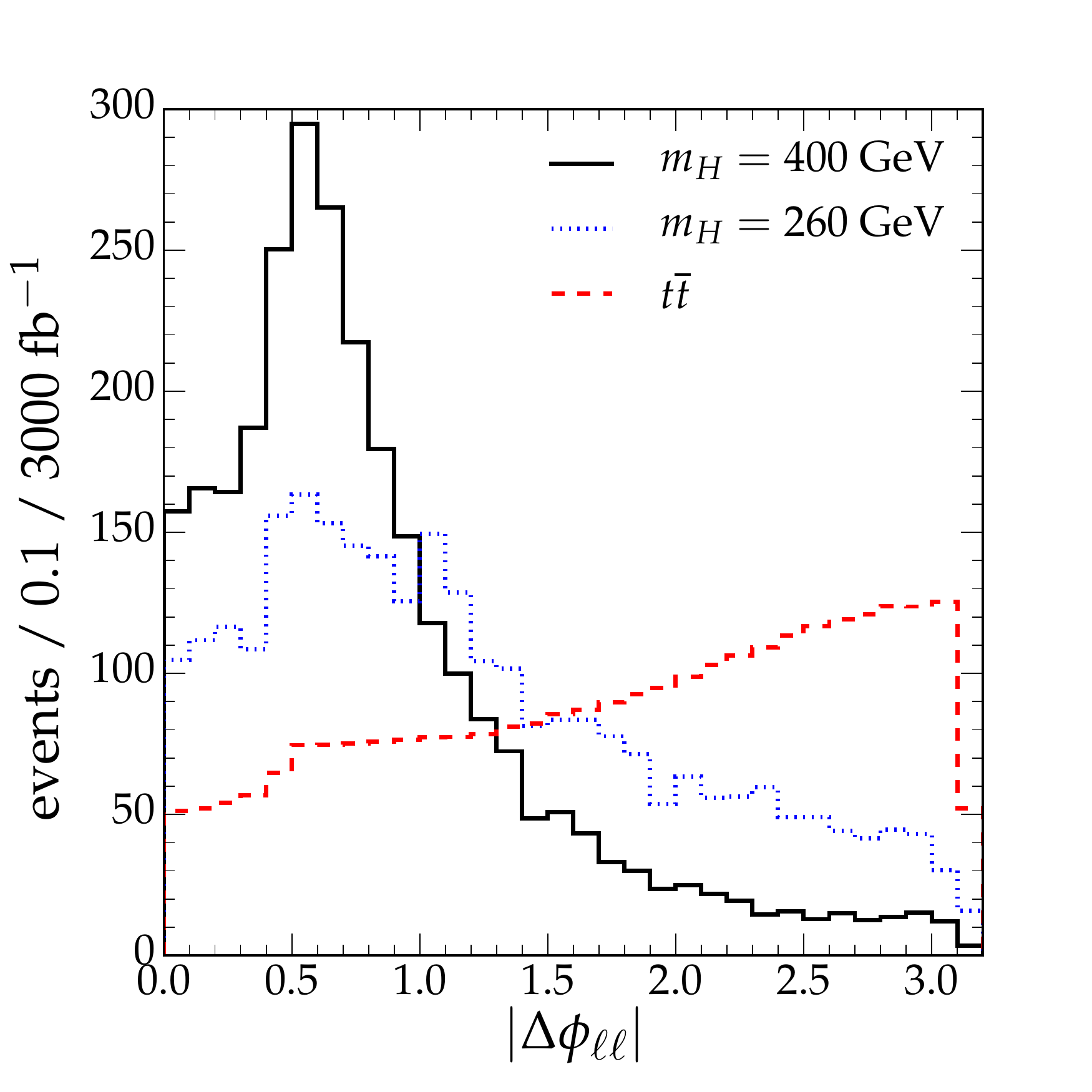}
    \includegraphics[width=0.48\textwidth]{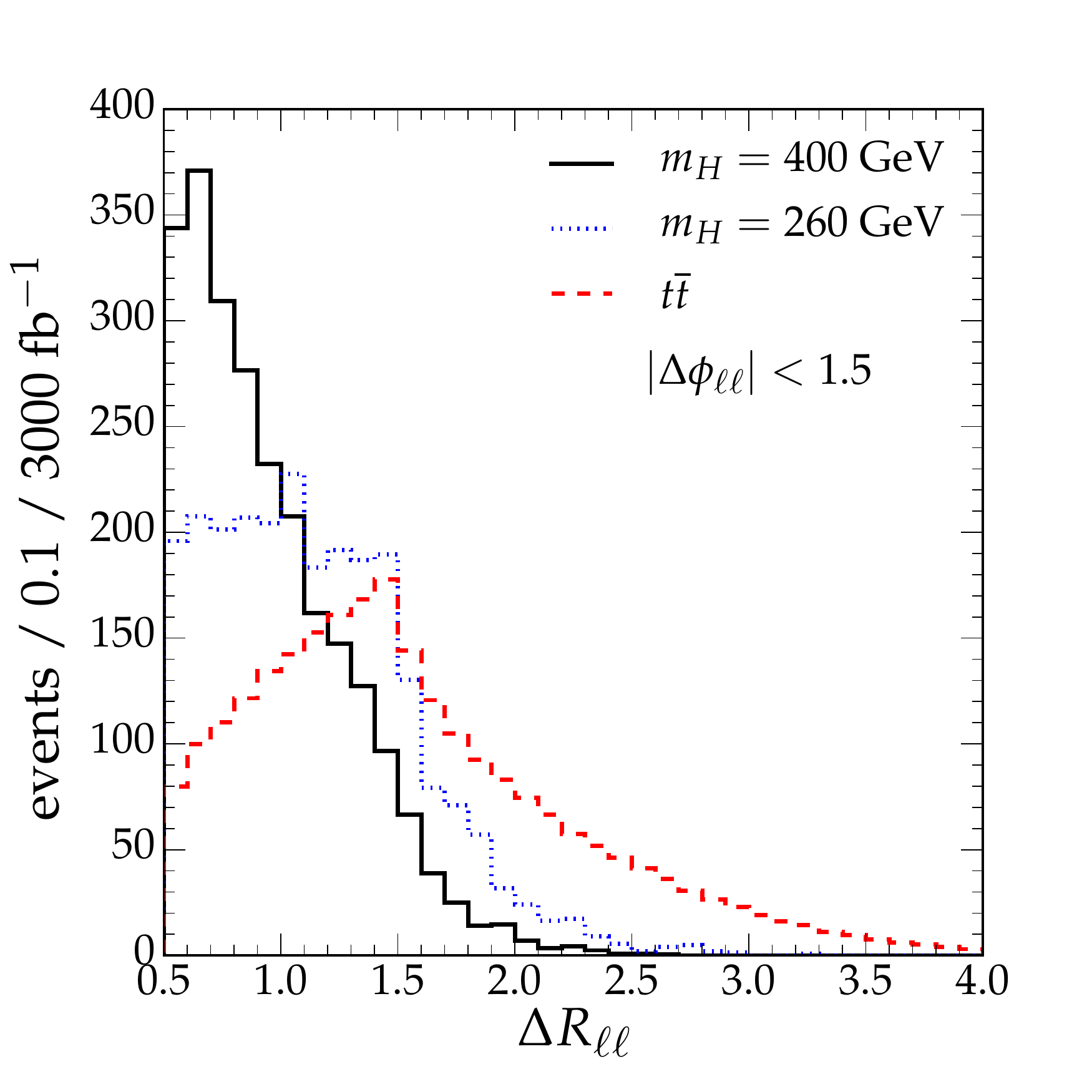}
    \includegraphics[width=0.48\textwidth]{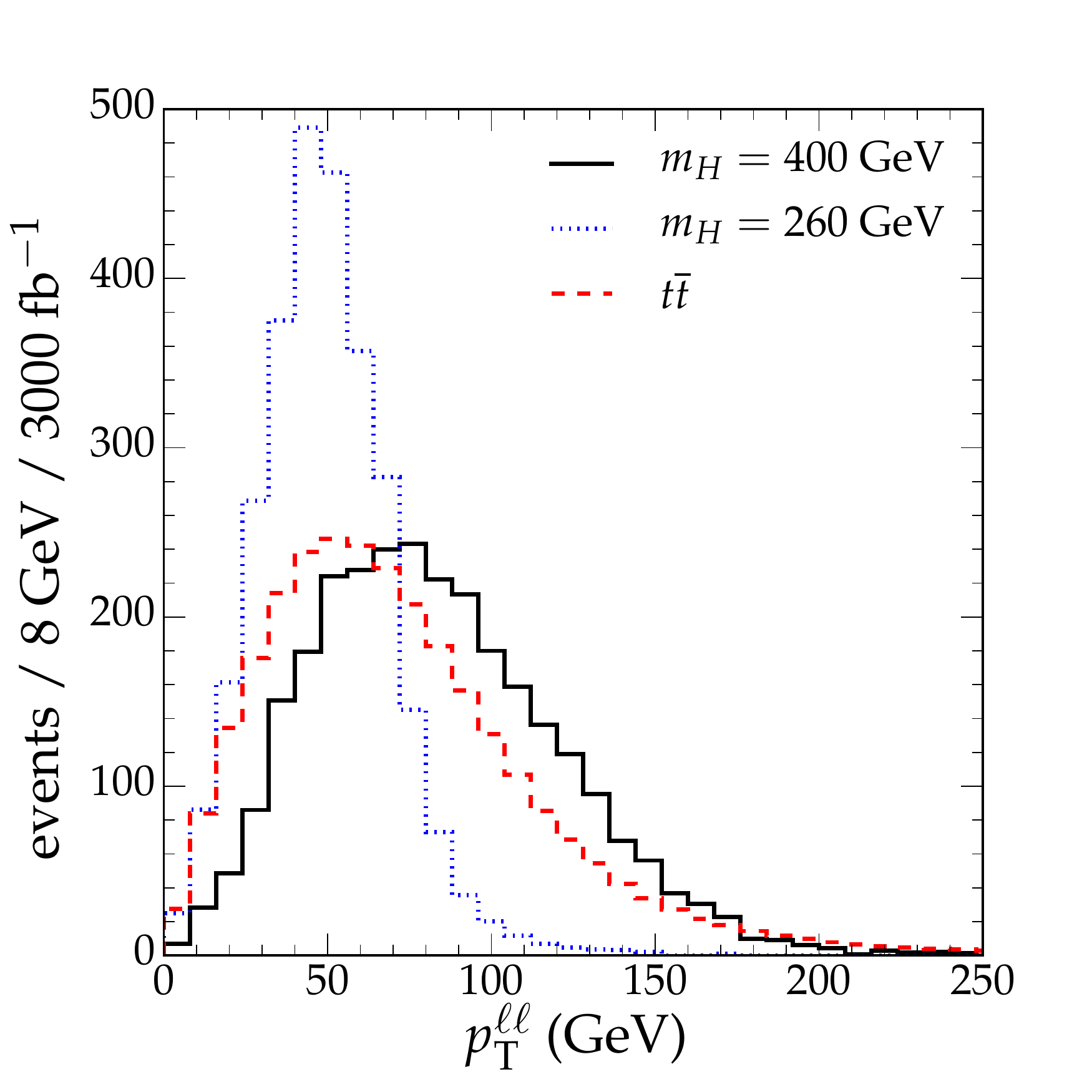}
    \includegraphics[width=0.48\textwidth]{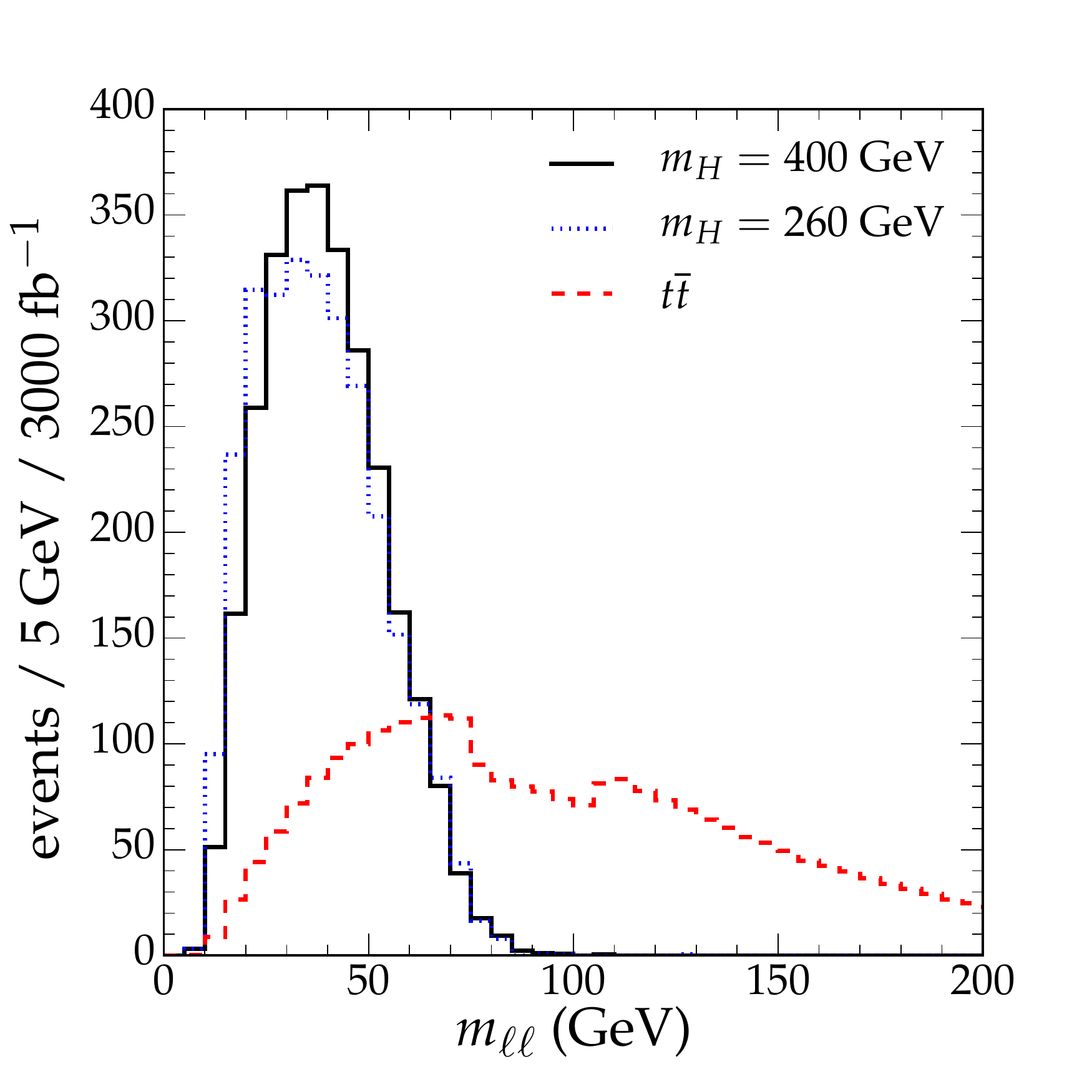}
  \end{center}
  \caption{Detector-level distributions of the kinematic variables for
    the two charged leptons. The upper frames are (Left panel) the
    azimuthal angular separations and (Right panel) the $\Delta
    R_{\ell\ell}$ when applying the azimuthal angular cut has been
    imposed. The lower frames are (Left panel) the sum of transverse
    momenta $p_{\rm T}^{\ell\ell}$ and (Right panel) the invariant
    mass $m_{\ell\ell}$ distributions. Basic
    selection cuts are applied and all the distributions are
    normalized for an illustration.}
\label{fig:dphi_ll}
\end{figure}
In the signal events, all the leptons are produced in the $h \to
WW^\ast$ decay process. In this case, it is known that the spin correlations
of the decay mode make the charged leptons collinear.
This feature can be used to further reduce the leptonic
backgrounds. We use two angular cuts: the azimuthal angular
difference $|\Delta \phi_{\ell\ell}| < 1.32$ (1.57) and $\Delta
R_{\ell\ell} \equiv \sqrt{(\Delta \phi_{\ell\ell})^2 + (\Delta
  \eta_{\ell\ell})^2} < 1.34$ (1.58) for the Higgs signal with
$m_H = 400$ (260) GeV.
The upper frames in Figure~\ref{fig:dphi_ll} show clear
separation between the signal and the $t\bar t$ background,
particularly when $m_H = 400$ GeV. This is because the
leptons can be much more boosted in the heavier Higgs events.
The collinearity of leptons is also encoded in the other cut
variables, the sum of the transverse momenta $p_{\rm
  T}^{\ell\ell} = |\mathbf{p}_{\rm T}^{\ell} + \mathbf{q}_{\rm T}^{\ell}|$
and the di-lepton invariant mass $m_{\ell\ell}$. In the case when $m_H =
260$ GeV, the leptons are less energetic so that the $p_{\rm
  T}^{\ell\ell}$ is relatively soft.
See the lower frames in Figure~\ref{fig:dphi_ll}. We require that
$p_{\rm T}^{\ell\ell} > 42$ (25) GeV and $m_{\ell\ell} < 60$ (47) GeV
for the $m_H = 400$ (260) GeV scenarios.
The $m_{\ell\ell}$ cut can also remove the $Z \to \tau^+\tau^-$
events in which the tau leptons decay leptonically.

\begin{figure}[t!]
  \begin{center}
    \includegraphics[width=0.48\textwidth]{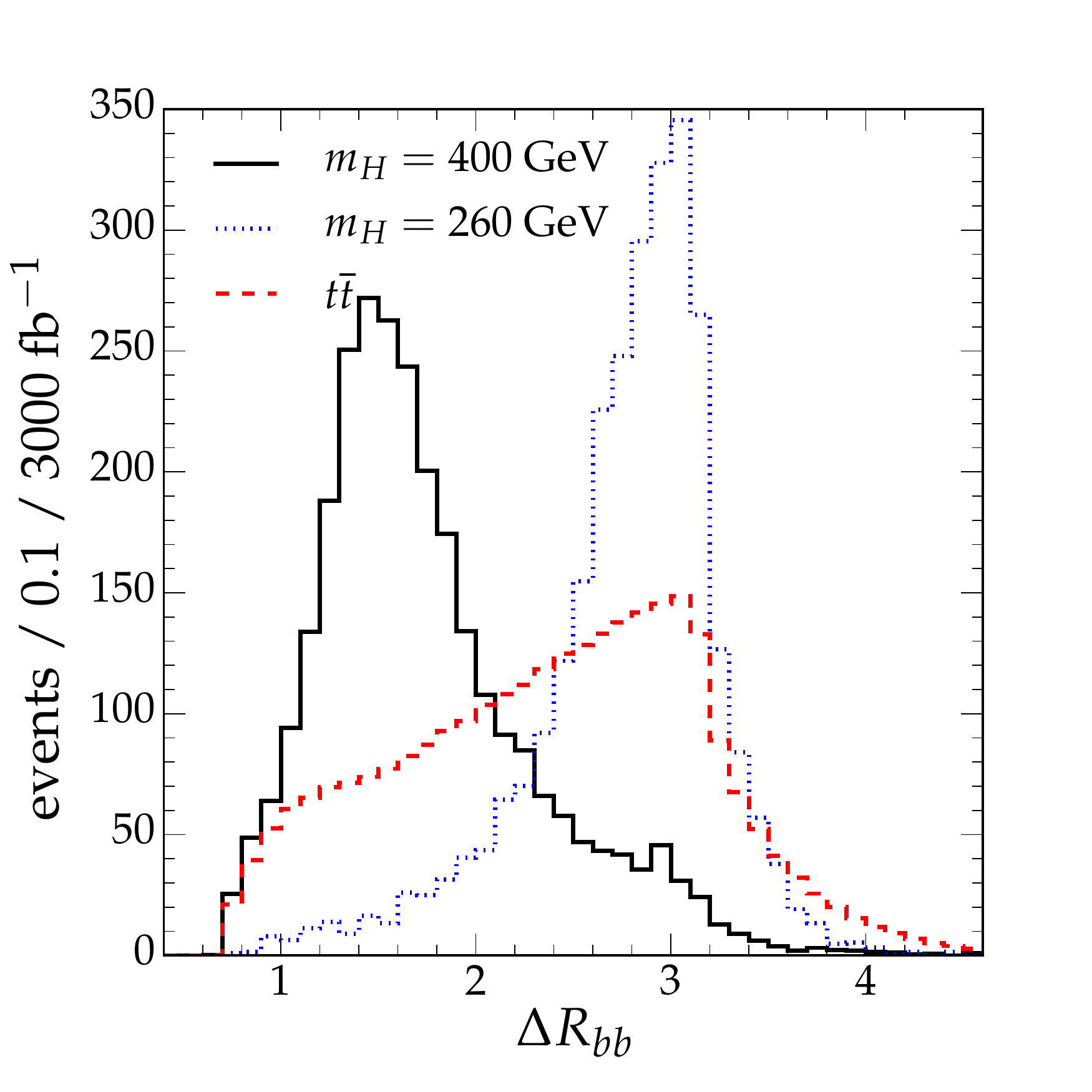}
    \includegraphics[width=0.48\textwidth]{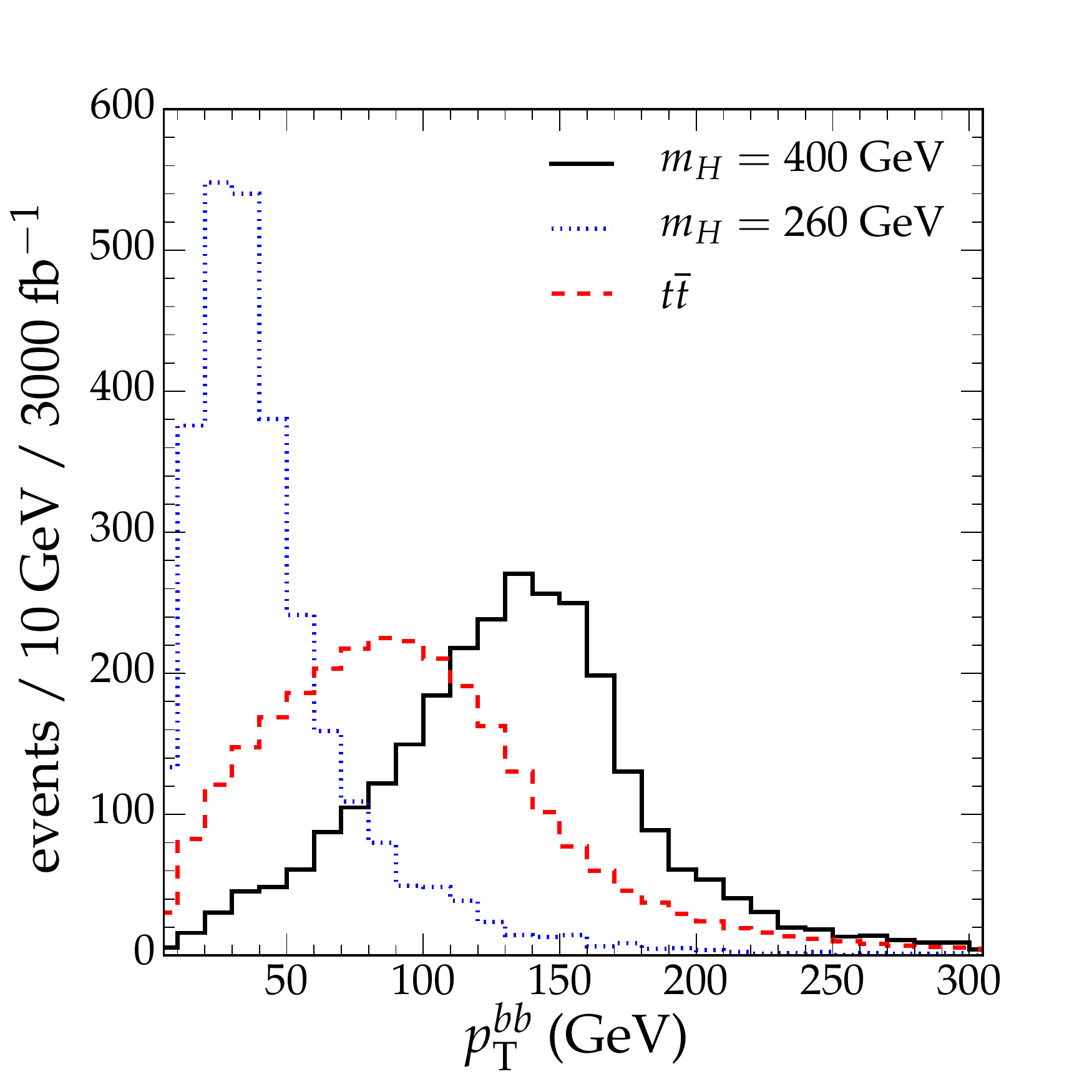}
    \includegraphics[width=0.48\textwidth]{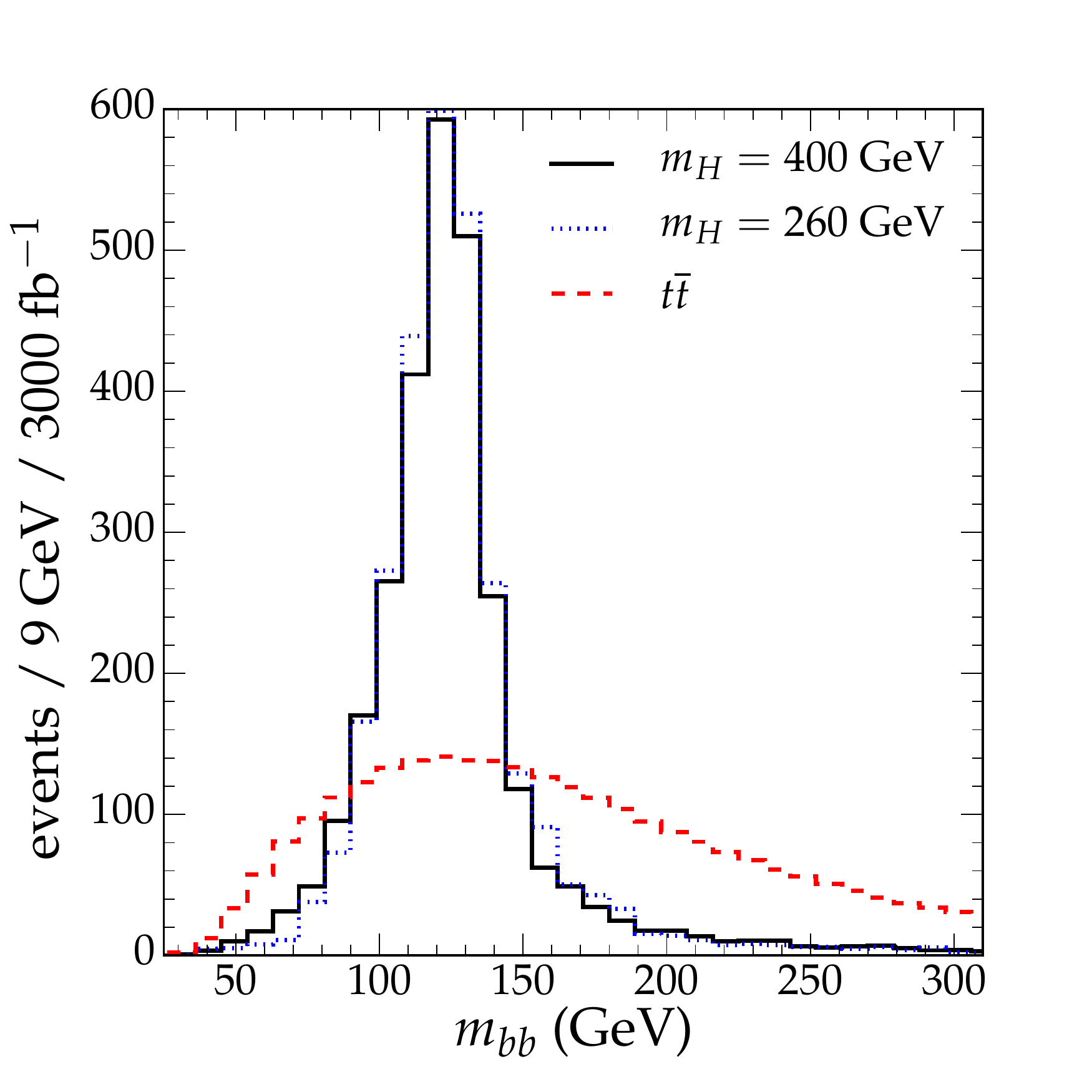}
    \includegraphics[width=0.48\textwidth]{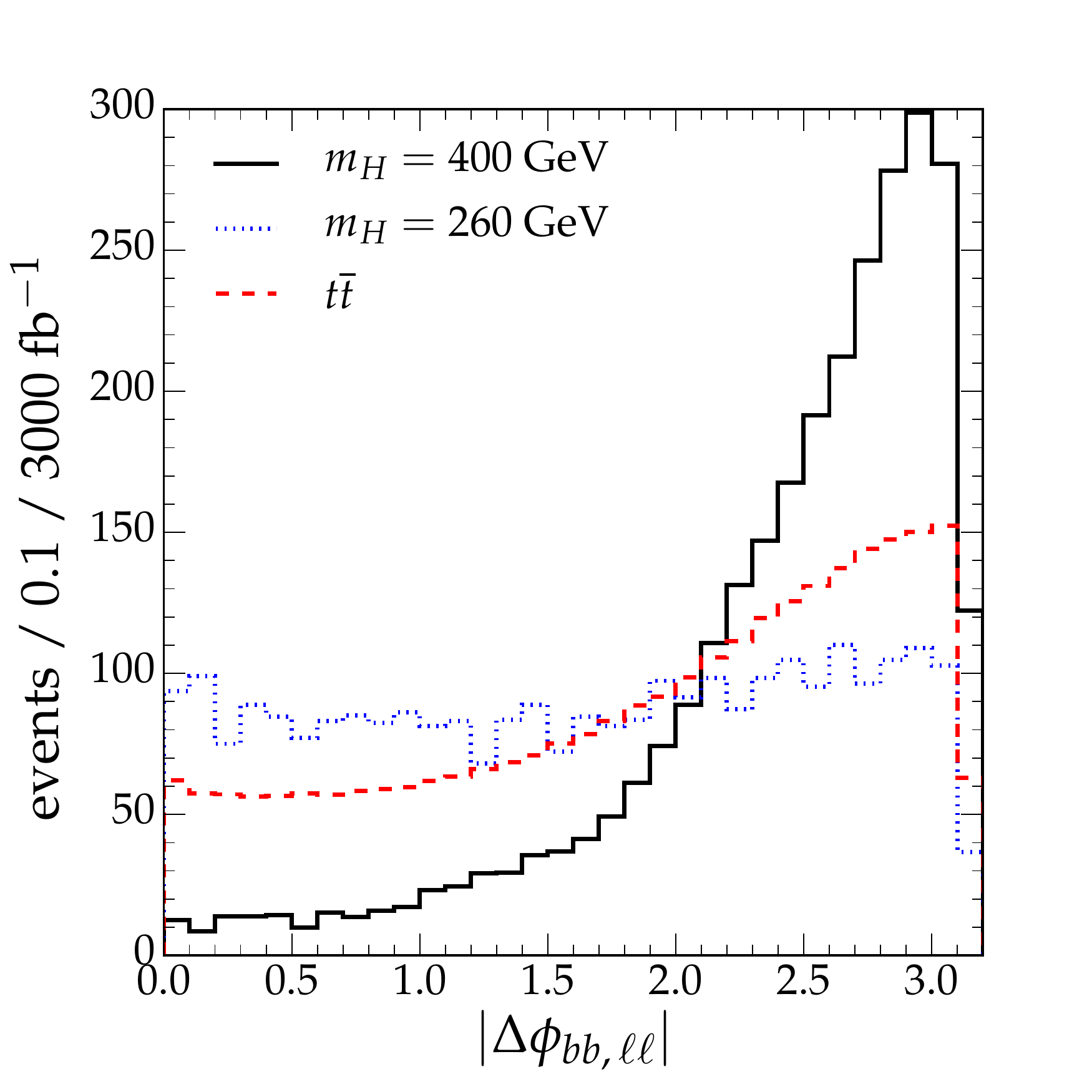}
  \end{center}
  \caption{Detector-level distributions of the kinematic variables for
    the two $b$-tagged jets. The upper frames are (Left panel) $\Delta
    R_{bb}$ and (Right panel) the transverse momentum $p_{\rm
      T}^{bb}$. The lower frames are (Left panel) the di-$b$-jet
    invariant mass and (Right panel) the azimuthal angular separation
    between $b\bar{b}$ and $\ell^+\ell^-$ systems. Basic
    selection cuts are applied and all the distributions are
    normalized for an illustration.}
\label{fig:dR_bb}
\end{figure}
In addition to the basic selection and the leptonic cuts, one can
impose cuts on the $b$-jets. Recently, a
boosted Higgs technique has been developed for processes like $pp
\to hV$ ($V = W$, $Z$)~\cite{Butterworth:2008iy} or $pp \to
hh$~\cite{Papaefstathiou:2012qe}, followed by $h \to
b\bar{b}$. In the situation where the Higgs boson is substantially
boosted, the jets produced by the Higgs boson can often be
considered as one fat jet, whose mass is around $m_h$.
For very high $p_{\rm T}^h \gg m_h$, $\Delta R_{bb} \equiv
\sqrt{(\Delta \phi_{bb})^2 + (\Delta \eta_{bb})^2}$ can be
estimated to be
\begin{align}
  \Delta R_{bb} \simeq \frac{2m_h}{p_T^h}.
  \label{eq:delta_R_bb}
\end{align}
If the fat Higgs jet condition could be applied, the backgrounds,
in particular, the $t\bar{t}$ events would be reduced very
efficiently since the $b$-jets in the background can have a relatively
large angular separation. In the Higgs signal, $p_{\rm T}^{h}$ can be
as large as
\begin{align}
  p_{\rm T}^h = \frac{m_H}{2}\sqrt{1 - \frac{4m_h^2}{m_H^2}} \simeq
  156~{\rm GeV}.
  \label{eq:pt_h}
\end{align}
so $\Delta R_{bb} \simeq 1.6$ for $m_H = 400$ GeV in the
rest frame of the heavy Higgs boson. The left panel in the upper frames
of Figure~\ref{fig:dR_bb} justifies this estimation.
Normally, the fat Higgs jet is required to have $\Delta R_{bb} \sim
1.2$ -- 1.5 or $p_{\rm T}^h \gtrsim 200$ GeV. Provided that the heavy
Higgs boson is produced at near-threshold energy, the transverse
momentum of the light Higgs has an upper bound as given
in~(\ref{eq:pt_h}). Therefore, we expect that the boosted Higgs
technique will be applicable in the case of much heavier Higgs boson
with $m_H \gtrsim 490$ GeV.

 In our benchmark points, it is inevitable to use the
conventional kinematic cuts. Although the angular separations of the
$b$-jets are rather sizable, it still turns out to be smaller than the
backgrounds in the case when $m_H = 400$ GeV, while the cut can
be applied in the opposite way in the case when $m_H = 260$ GeV.
This can be easily deduced from eq.~(\ref{eq:delta_R_bb}), which
predicts that the angular separation can be very large for the smaller
$p_{\rm T}^{h}$ value.
On the other hand, the right panel in the upper frames of
Figure~\ref{fig:dR_bb} shows that $m_H = 400$ GeV signal events
possess much larger values of the total transverse momentum for the
$b\bar{b}$ system.
We select events with $\Delta R_{bb} < 2.25$ and $p_{\rm
  T}^{bb} > 105$ GeV for $m_H = 400$ GeV, while $\Delta R_{bb} > 2.56$
without imposing any $p_{\rm T}^{bb}$ cut for $m_H = 260$ GeV signal
events.
Since the $m_h$ value is already known, one can further impose a cut
on the di-$b$-jet invariant mass to ensure that the $b$-jets are
originated from the light Higgs boson.
One can see that the invariant mass distributions have clear peaks
around $m_h = 125$ GeV for both benchmark points in the left panel in
the lower frames of Figure~\ref{fig:dR_bb}.
The invariant mass is required to have a value within a mass window
of $115~(94)~{\rm GeV} < m_{bb}< 146$ (135) GeV for $m_H = 400$ (260)
GeV signal.

In the case when the heavy Higgs boson is produced  near threshold,
the light Higgs boson pair will be almost in a back-to-back
configuration. Then, it is likely that the direction of the $b\bar{b}$
system will be well separated from that of the $\ell^+\ell^-$
system. This feature can be seen in the right panel in the lower frames
of Figure~\ref{fig:dR_bb}, where the distributions for the absolute value of
$\Delta \phi_{bb,\,\ell\ell} \equiv
\cos^{-1}(\mathbf{\hat{p}}_{\rm T}^{bb} \cdot
\mathbf{\hat{p}}_{\rm T}^{\ell\ell})$, where $\mathbf{\hat{p}}_{\rm T}
\equiv \mathbf{p}_{\rm T} / p_{\rm T}$, are shown. We take events with
$|\Delta \phi_{bb,\,\ell\ell}| > 1.92$ for $m_H = 400$ GeV
signal. This cut is not applicable to the case of the $m_H = 260$ GeV
signal as the angular separation can be relatively small due to the small
boost of each Higgs decay chain.

\begin{figure}[t!]
  \begin{center}
    \includegraphics[width=0.48\textwidth]{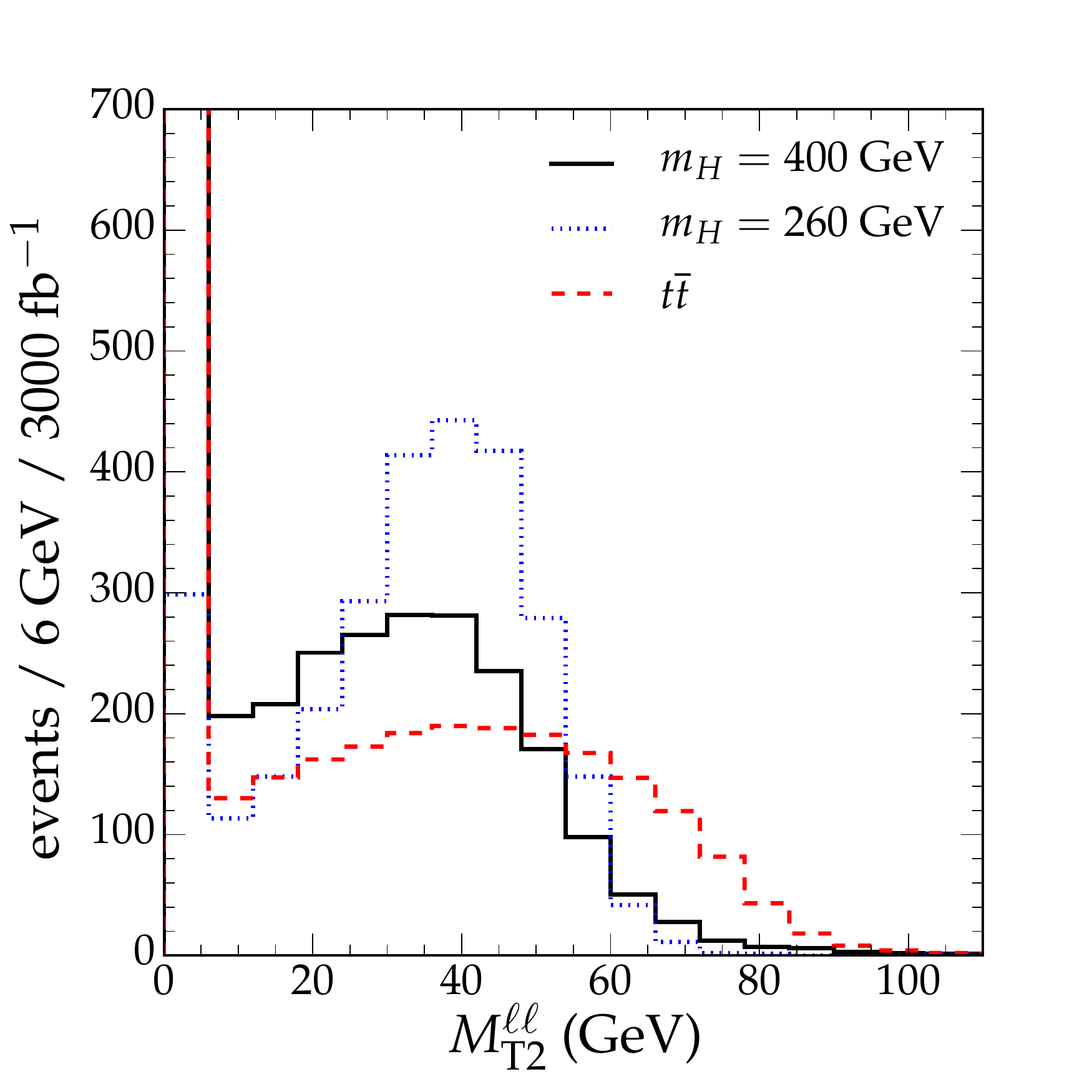}
    \includegraphics[width=0.48\textwidth]{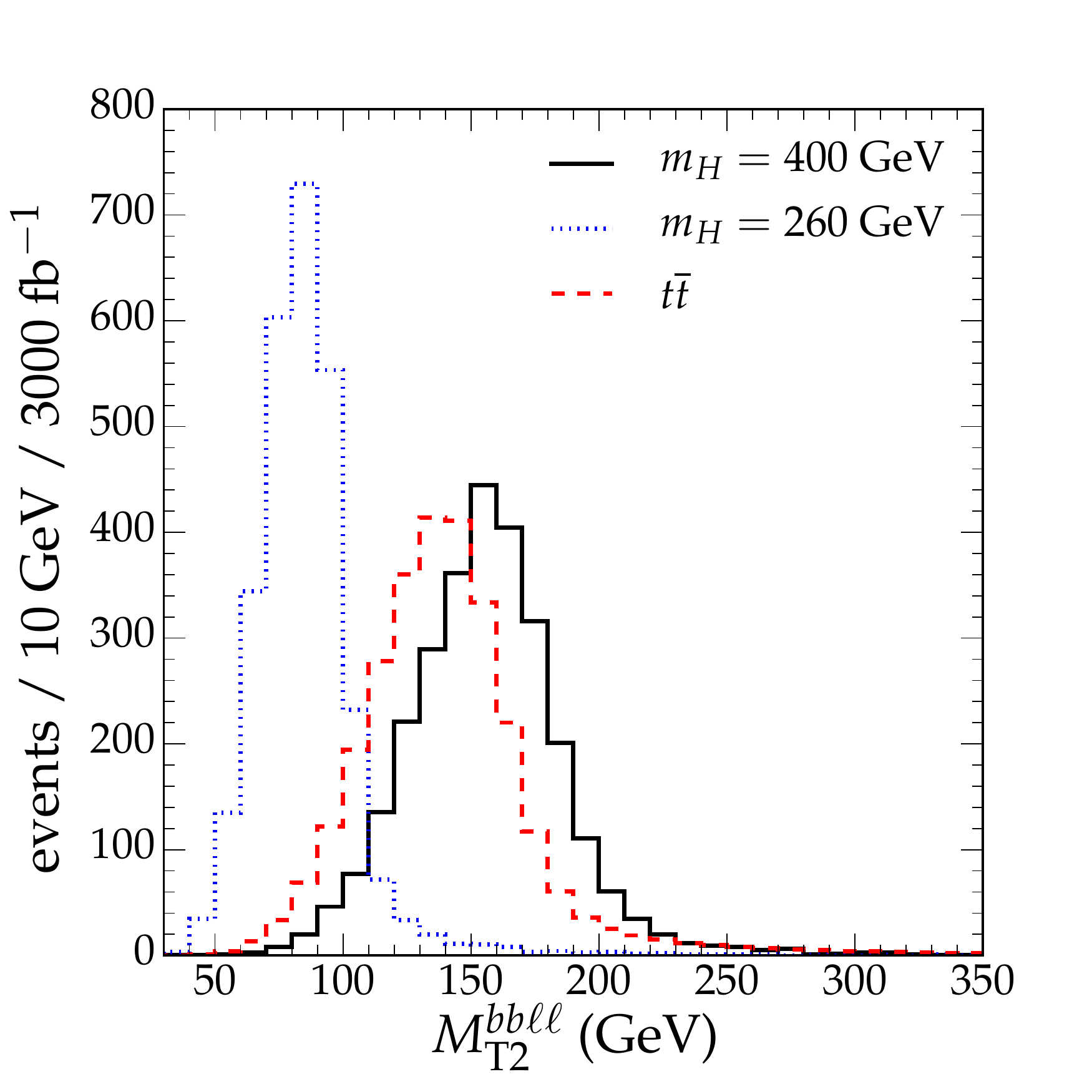}
    \includegraphics[width=0.48\textwidth]{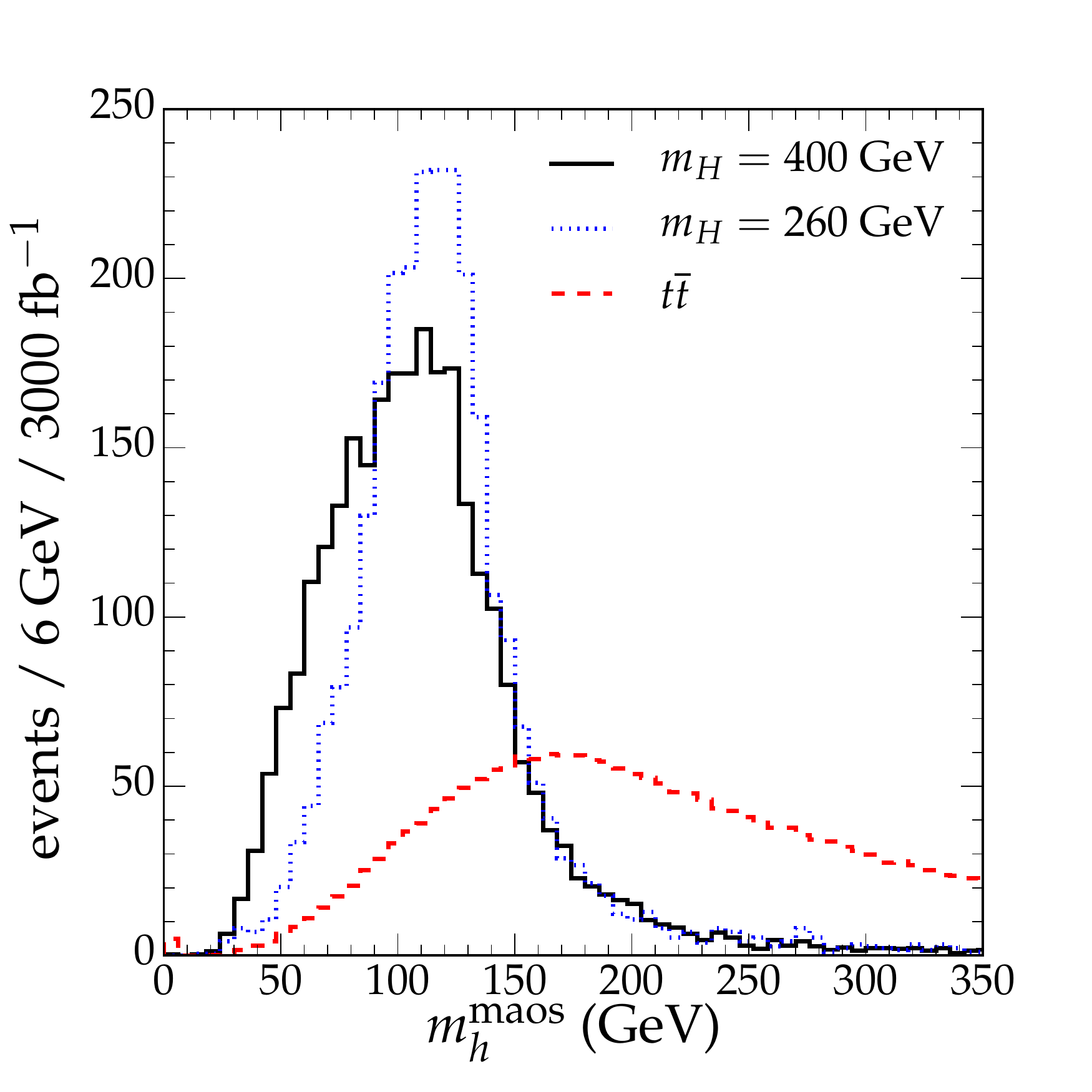}
    \includegraphics[width=0.48\textwidth]{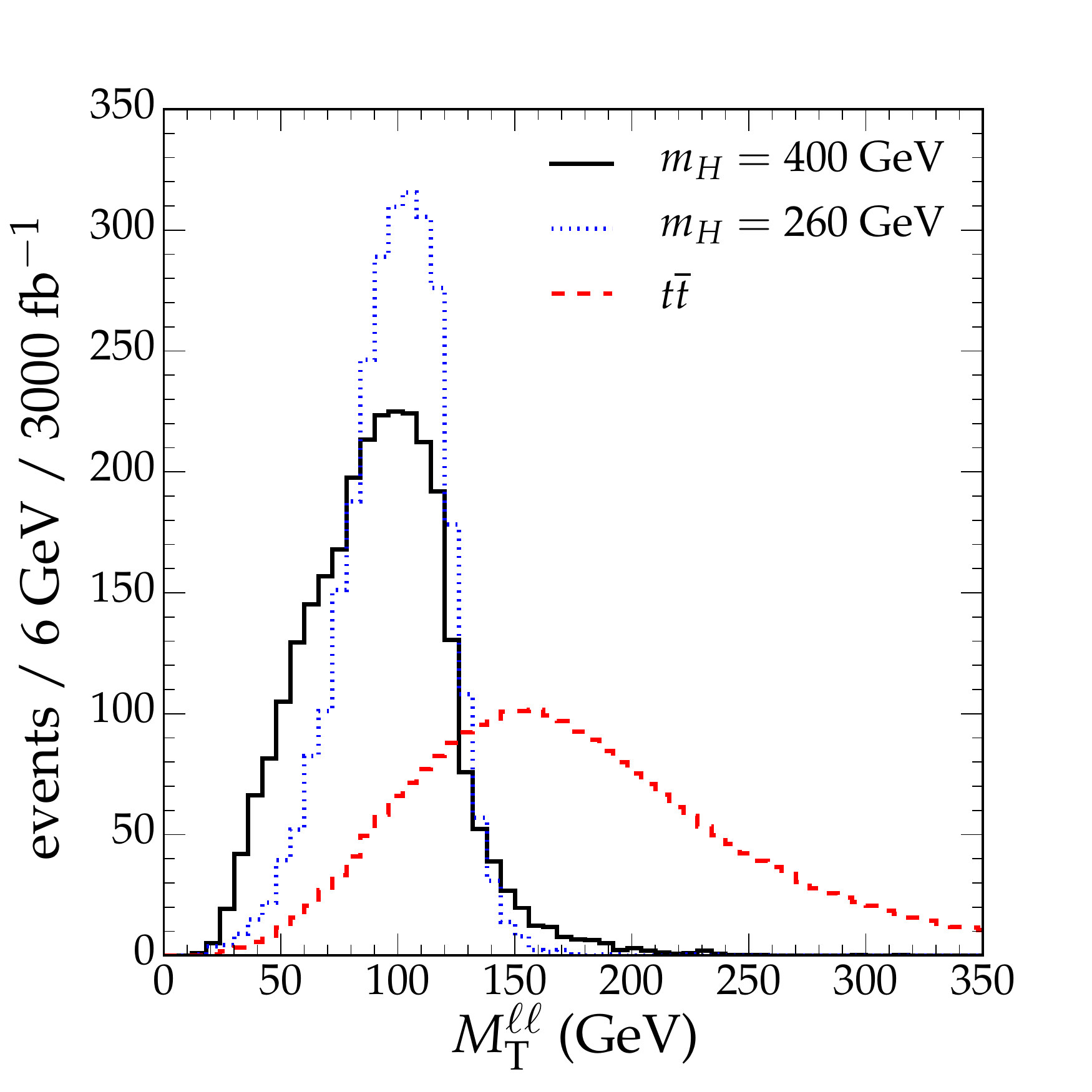}
  \end{center}
  \caption{The upper frames are detector-level $M_{\rm T2}$
    distributions for (Left panel)
    the $2\ell + \slashed{E}_{\rm T}$ and (Right panel) $2b + 2\ell +
    \slashed{E}_{\rm T}$ systems. The lower frame are (Left panel)
    $m_h^{\rm maos}$ and (Right panel) $M_{\rm T}^{\ell\ell}$
    distributions for detector-level signals and backgrounds. Basic
    selection cuts are applied and all the distributions are
    normalized for an illustration.}
\label{fig:m_T2}
\end{figure}

We now turn to the $M_{\rm T2}$ cuts. For the $2b + 2\ell + \slashed{E}_{\rm T}$ final state, one can construct two kinds of
$M_{\rm T2}$ according to the definition for the visible $+$ invisible
system, that is, either $2\ell + \slashed{E}_{\rm T}$, which contains
leptons only, or $2b + 2\ell + \slashed{E}_{\rm T}$, which contains
$b$-jets as well as leptons when forming the visible particle system.
We emphasize that $M_{\rm T2}$ has been known to be applicable to
a system that can be divided into two groups of visible particles like
processes depicted in the decay topology
(\ref{eq:pair_decay_topology}) with a pair production of heavy
particles, followed by two separate decay chains.
The $2\ell + \slashed{E}_{\rm T}$ system in the signal decay topology
(\ref{eq:signal_topology}) can be regarded as such a process.
In what follows, the $M_{\rm T2}$
calculated for the $2\ell + \slashed{E}_{\rm T}$ system is expressed as
$M_{\rm T2}^{\ell\ell}$ to distinguish it from the other kind of
$M_{\rm T2}$.
As is derived in Appendix~\ref{sec:m_T2_Higgs} for some kinematic
configurations, the $M_{\rm T2}^{\ell\ell}$ distribution is bounded
from above by $m_h / 2 < m_W$, whereas it has a maximum at $m_W$ in
the di-leptonic $t\bar{t}$ process since both $W$ bosons are in
on-shell.
The $M_{\rm T2}^{\ell\ell}$ distributions in the left panel of
Figure~\ref{fig:m_T2} clearly show the endpoint structure.
Another notable feature is that there are a number of events
which have vanishing $M_{\rm T2}^{\ell\ell}$ for both signal and
background distributions. It corresponds to a trivial zero of $M_{\rm
  T2}^{\ell\ell}$ in the fully massless case, \ie, $m_\ell = m_\nu =
0$~\cite{Lester:2011nj}. This happens when the missing transverse
momentum $\slashed{\mathbf{p}}_{\rm T}$ lies on the smaller
sector of the transverse plane spanned by the visible momentum vectors
$\mathbf{p}_{\rm T}^{\ell}$ and $\mathbf{q}_{\rm T}^{\ell}$. In this
case, the $M_{\rm T2}^{\ell\ell}$ value is taken for a momentum
partition where both transverse masses in eqs.~(\ref{eq:m_T}) are
vanishing.
However, the fraction of events with the trivial zero of the
$M_{\rm T2}^{\ell\ell}$ can be different depending on the preferred
momentum configuration of the process. Due to the spin
correlation and the boost, the opening angle of the charged
leptons in the Higgs signal event is smaller than that in the
di-leptonic $t\bar{t}$ events as was seen in the upper frames of
Figure~\ref{fig:dphi_ll}. It means that there are more chances to have
the trivial zero of $M_{\rm T2}^{\ell\ell}$ in the $t\bar{t}$ events
than in the Higgs signal. Therefore, the lower cut as well as the upper
one can help reduce the backgrounds further. This lower cut on the
$M_{\rm T2}^{\ell\ell}$ also increases the accuracy of the MAOS
momenta, which will be used in the subsequent analysis. We impose the
$M_{\rm T2}^{\ell\ell}$ cut as $25~{\rm GeV} < M_{\rm T2}^{\ell\ell} <
60$ GeV for $m_H = 260$ GeV signal. In the case when $m_H = 400$ GeV,
the missing transverse momentum vector can lie
inside of the opening angle of the di-lepton system when the light
Higgs is fairly boosted. Therefore, we do not apply the lower cut, but
only the upper cut $M_{\rm T2}^{\ell\ell} < 60$ GeV is imposed for the
$m_H = 400$ GeV signal.

Once $M_{\rm T2}^{\ell\ell}$ has been calculated, one can
construct the invariant mass of the $2\ell + \slashed{E}_{\rm T}$ system
by using the MAOS momentum of the invisible particle given as
\begin{align}
  \left ( m_h^{\rm maos} \right )^2 \equiv \left ( p^\ell + q^\ell +
    k^{\rm maos} + l^{\rm maos} \right )^2 ,
\end{align}
which equals to $m_h$ when $k^{\rm maos} = k^{\rm true}$ and
$l^{\rm maos} = l^{\rm true}$. It is shown in the left panel in the
lower frame of Figure~\ref{fig:m_T2}. One can further employ the
transverse mass of the leptonic system while ignoring the unknown
$m_{\nu\nu}$ value and the longitudinal momentum components of
neutrinos,
\begin{align}
  \left ( M_{\rm T}^{\ell\ell} \right )^2 = m_{\ell\ell}^2 +
  2 \left (\sqrt{|\mathbf{p}_{\rm T}^{\ell\ell}|^2 +
    m_{\ell\ell}^2}|\slashed{\mathbf{p}}_{\rm T}| - \mathbf{p}_{\rm
    T}^{\ell\ell} \cdot \slashed{\mathbf{p}}_{\rm T} \right ) ,
\end{align}
which is bounded from above by $m_h$ \cite{Barr:2009mx}.
Since both distributions have
distinguishable peak and edge structures as well as a strong correlation
with $m_h$, we use them as cut variables demanding $m_{h}^{\rm maos} <
145$ GeV and $30~{\rm GeV} < M_{\rm T}^{\ell\ell} < 125$ GeV for $m_H
= 400$ GeV, and $60~{\rm GeV} < m_h^{\rm maos} < 136$ GeV and $58~{\rm
GeV} < M_{\rm T}^{\ell\ell} < 126$ GeV for $m_H = 260$ GeV signal
events. We have not applied the lower cut on $m_h^{\rm maos}$ for
$m_H = 400$ GeV since the distribution is relatively distorted due to
the trivial zero solutions described above.

After counting two $b$-jets as well as the charged leptons among
the set of visible particle system, \ie, $V = b\ell$,
one can define another kind of the $M_{\rm T2}$ variable, denoted as
$M_{\rm T2}^{bb\ell\ell}$.\footnote{
  There is an ambiguity of how to pair one $b$-jet to one charged
  lepton because there can be two possible pairings in each event.
  Here, we define $M_{\rm T2}^{bb\ell\ell}$ as the
  smaller one between two possible values of $M_{\rm
    T2}^{bb\ell\ell}$. This definition matches the one used to measure
  the top quark mass using $M_{\rm T2}$ in~\cite{Cho:2008cu}.
}
Recall that $M_{\rm T2}$ aims at the physics processes describable
by the decay topology (\ref{eq:pair_decay_topology}).
The Higgs signal has a different decay topology since
the invisible particle system is disjointed from the $b\bar{b}$
system.
On the other hand, it is well known that the di-leptonic
$t\bar{t}$ process is the one of the SM processes where the $M_{\rm
  T2}$ variable is applicable since the decay topology is exactly the
same as (\ref{eq:pair_decay_topology}), and the $M_{\rm
  T2}^{bb\ell\ell}$ distribution is strictly bounded from
above by $m_t$. Therefore, one can still attempt to employ $M_{\rm
  T2}^{bb\ell\ell}$ to reduce the $t\bar{t}$ backgrounds if the edge
structure of the signal distribution has a certain amount of deviation
from $m_t$. The $M_{\rm T2}^{bb\ell\ell}$ distributions for both
signal and $t\bar{t}$ are shown in the right panel of
Figure~\ref{fig:m_T2}.

Before going further, we here briefly summarize the types of the
$M_{\rm T2}$ solutions for the invisible momenta.
The hypothetical invisible momentum configuration that gives the $M_{\rm
  T2}$ value can be classified in two types. One is a balanced
configuration, in which $M_{\rm T}^{(1)} = M_{\rm T}^{(2)}$ is
realized, and the other is an unbalanced one, in which $M_{\rm T}^{(1)}
\neq M_{\rm T}^{(2)}$~\cite{Barr:2003rg}.
In each collider event, only one type
of the momentum configuration provides the $M_{\rm T2}$ value, and it
can be deduced by the invariant masses of the visible particle set in
the event, $m_{V}$ and $m_{\overline{V}}$ in
eqs.~(\ref{eq:m_T}). One can easily find that a stationary value of
the transverse mass $M_{\rm T}^{(1)}$ is attained when
$\mathbf{k}_{\rm T} = m_\chi \mathbf{p}_{\rm T} / m_{V}$ and
$\mathbf{l}_{\rm T} = \slashed{\mathbf{p}}_{\rm T} - \mathbf{k}_{\rm
  T}$. Then,
\begin{align}
  M_{\rm T}^{(1)} = m_{V} + m_\chi ,
\end{align}
which is called an unconstrained minimum of the transverse mass.
Similarly, one can find the stationary value of $M_{\rm T}^{(2)} =
m_{\overline{V}} + m_\chi$. For each stationary point, the
$M_{\rm T}^{(1)}$ value can be compared to $M_{\rm
  T}^{(2)}$. In the case that
\begin{align}
  \left . M_{\rm T}^{(1)} \right |_{\mathbf{k}_{\rm T} = m_\chi
    \mathbf{p}_{\rm T} / m_{V}} = m_{V} + m_\chi >
  \left . M_{\rm T}^{(2)} \right |_{\mathbf{l}_{\rm T} =
    \slashed{\mathbf{p}}_{\rm T} - \mathbf{k}_{\rm T}},
\end{align}
$M_{\rm T2}$ is given by the unconstrained minimum of $M_{\rm
  T}^{(1)}$, \ie,
\begin{align}
  M_{\rm T2} = m_{V} + m_\chi .
  \label{eq:m_T2_unbal}
\end{align}
This corresponds to the unbalanced configuration. On the other hand,
if it is satisfied that
\begin{align}
  \left . M_{\rm T}^{(1)} \right |_{\mathbf{k}_{\rm T} = m_\chi
    \mathbf{p}_{\rm T} / m_{V}} = m_{V} + m_\chi &
  \leq
  \left . M_{\rm T}^{(2)} \right |_{\mathbf{l}_{\rm T} =
    \slashed{\mathbf{p}}_{\rm T} - \mathbf{k}_{\rm T}}, \nonumber\\
  \left . M_{\rm T}^{(2)} \right |_{\mathbf{l}_{\rm T} = m_\chi
    \mathbf{q}_{\rm T} / m_{\overline{V}}} =
  m_{\overline{V}} + m_\chi &
  \leq
  \left . M_{\rm T}^{(1)} \right |_{\mathbf{k}_{\rm T} =
    \slashed{\mathbf{p}}_{\rm T} - \mathbf{l}_{\rm T}},
\end{align}
then $M_{\rm T2}$ is given by the balanced configuration in which
$M_{\rm T}^{(1)} = M_{\rm T}^{(2)}$.
See refs.~\cite{Barr:2003rg,  Lester:2007fq, Cho:2007qv, Cho:2007dh}
for the detailed discussion of the momentum configuration types and
their corresponding properties of $M_{\rm T2}$.

\begin{figure}[t!]
  \begin{center}
    \includegraphics[width=0.48\textwidth]{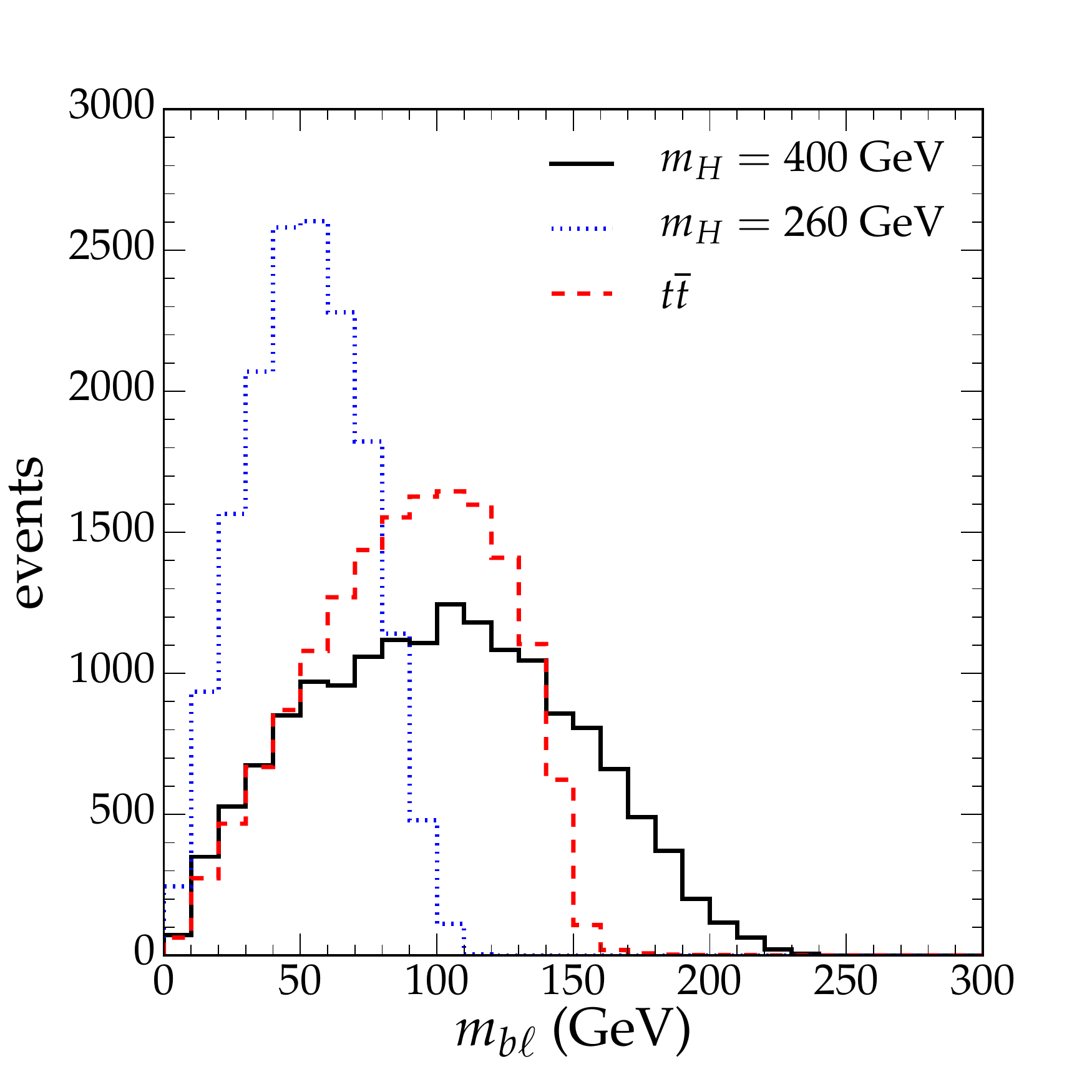}
    \includegraphics[width=0.48\textwidth]{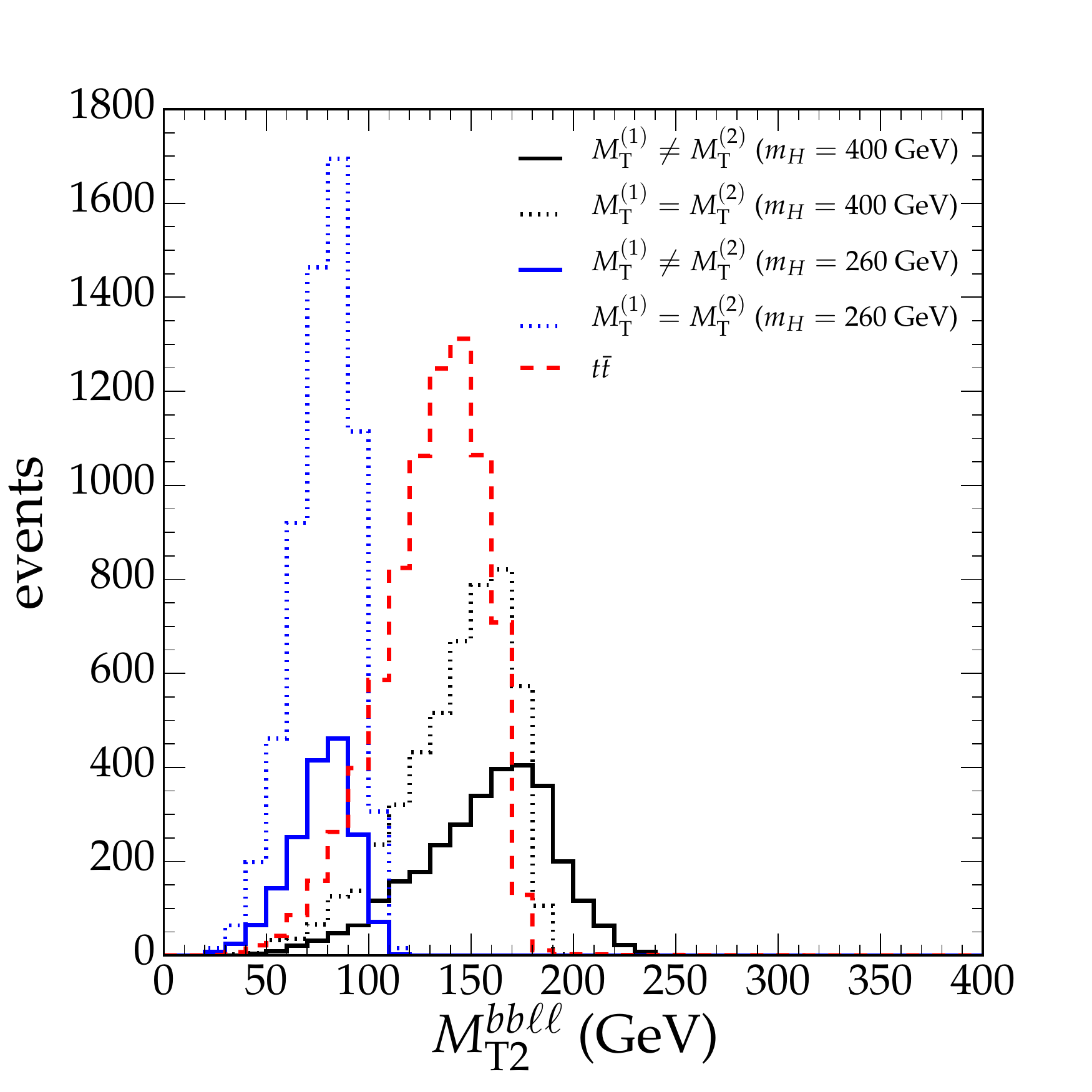}
  \end{center}
  \caption{Parton-level distributions for (Left panel) $m_{b\ell}$
    when $m_H = 260$ and 400 GeV and (Right panel) $M_{\rm
      T2}^{bb\ell\ell}$ classified by the types of the invisible
    momentum configuration chosen by the $M_{\rm T2}$ calculation. See
    the text for detailed explanation. For a comparison,
    distributions for the di-leptonic $t\bar{t}$ process are shown.}
\label{fig:m_bl}
\end{figure}

In the case of $M_{\rm T2}^{\ell\ell}$, the $M_{\rm T2}$ value is always
given by the balanced configuration since $m_\ell = m_\nu =
0$. On the other hand, because $m_{b\ell}$ is not a constant but a
variable, there exist sort of events in which the unbalanced
configuration is selected to provide the $M_{\rm T2}^{bb\ell\ell}$
value. In the di-leptonic $t\bar{t}$ process,
\begin{align}
  m_{b\ell} \leq \sqrt{m_t^2 - m_W^2} \simeq 154~{\rm GeV}
  \label{eq:m_bl_max_tt}
\end{align}
when the $b$ quark mass is neglected.
Therefore, the unbalanced $M_{\rm T2}^{bb\ell\ell}$ has a maximum
value smaller than $m_t$, while the balanced $M_{\rm T2}$ value can be
as large as $m_t$. This means that the overall $M_{\rm T2}^{bb\ell\ell}$
distribution is bounded from above by the maximum of the balanced $M_{\rm
  T2}^{bb\ell\ell}$ values.
For the Higgs signal, the situation is different. If one considers the
case when the total transverse momentum of whole system is vanishing,
or identically, the heavy Higgs has been produced at rest on the
transverse plane, one can find that the balanced $M_{\rm
  T2}^{bb\ell\ell}$ value cannot exceed $m_H / 2$ by a similar
consideration as done in Appendix~\ref{sec:m_T2_Higgs}. However, by
eq.~(\ref{eq:m_T2_unbal}),
the unbalanced $M_{\rm T2}^{bb\ell\ell}$ has an upper bound at
$m_{b\ell}^{\rm max}$, which can be expressed analytically as
\begin{align}
  m_{b\ell}^{\rm max} = \frac{m_H m_W}{2m_h} \left ( 1 +
    \sqrt{1 - \frac{4 m_h^2}{m_H^2}} \right ) \simeq 229~{\rm GeV}
  \label{eq:m_bl_max}
\end{align}
for $m_H = 400$ GeV, while it is $\simeq 107$ GeV for $m_H = 260$ GeV.
The maximum value in the above equation is achieved when
one of the hypothetical neutrino momenta chosen by the $M_{\rm
  T2}^{bb\ell\ell}$ calculation is parallel to the momentum direction
of the charged lepton sharing the same parent particle, while the
other one is anti-parallel.\footnote{We note that this $m_{b\ell}^{\rm max}$
for the Higgs signal is not a global maximum for all possible $b\ell$
pairings, but the maximum for a pair which leads to the smaller value
of $M_{\rm T2}^{bb\ell\ell}$.} %
The $m_{b\ell}$ distributions for various $m_H$
values and the $M_{\rm T2}^{bb\ell\ell}$ distributions classified by
the types of the $M_{\rm T2}^{bb\ell\ell}$ solutions
are shown in Figure~\ref{fig:m_bl}, using the parton-level data for
the sake of a numerical demonstration.
This also means that the endpoint of $M_{\rm T2}^{bb\ell\ell}$ as
well as the $m_{b\ell}$ distributions for the Higgs signal events will
be smaller than $m_t$ if $m_H \lesssim 330$ GeV, and in that case,
the upper cut instead of the lower one should be used unless
the upper bound value is too close to $m_t$.
This observation may lead one to deduce that the efficiency of the
$M_{\rm T2}^{bb\ell\ell}$ cut might be the similar as that
of the $m_{b\ell}$ cut. However, in our numerical study, the $M_{\rm
  T2}^{bb\ell\ell}$ cut turns out to perform slightly better than
$m_{b\ell}$.
This might be because $M_{\rm T2}^{bb\ell\ell}$ incorporates the
effect of the missing momentum and its correlation with the visible
momenta. We set the event selection cut value as $M_{\rm
  T2}^{bb\ell\ell} > 165$ GeV for $m_H = 400$ GeV and $M_{\rm
  T2}^{bb\ell\ell} < 96$ GeV for $m_H = 260$ GeV signals.

When considering final-state particles all together, the simplest
kinematic variables that one can construct are the invariant mass of
total visible system, $m_{bb\ell\ell}$, and the transverse mass of the
full system including the missing energy. Since the full visible $+$
invisible system has a fixed invariant mass, \ie, $m_H$, the
invariant mass of the visible system also has a dependency on $m_H$ for
its maximal value. One can find that
\begin{align}
  m_{bb\ell\ell}^2 \leq \frac{m_H^2}{2} \left (1 + \sqrt{1 -
      \frac{4m_h^2}{m_H^2}}
    \right ) \simeq \left ( 377~{\rm GeV} \right )^2
\end{align}
for $m_H = 400$ GeV, whereas there is no definite cut-off in the
$t\bar{t}$ background since $m_{t\bar{t}}$ is a variable of the event
in the hadron collider.
\begin{figure}[t!]
  \begin{center}
    \includegraphics[width=0.48\textwidth]{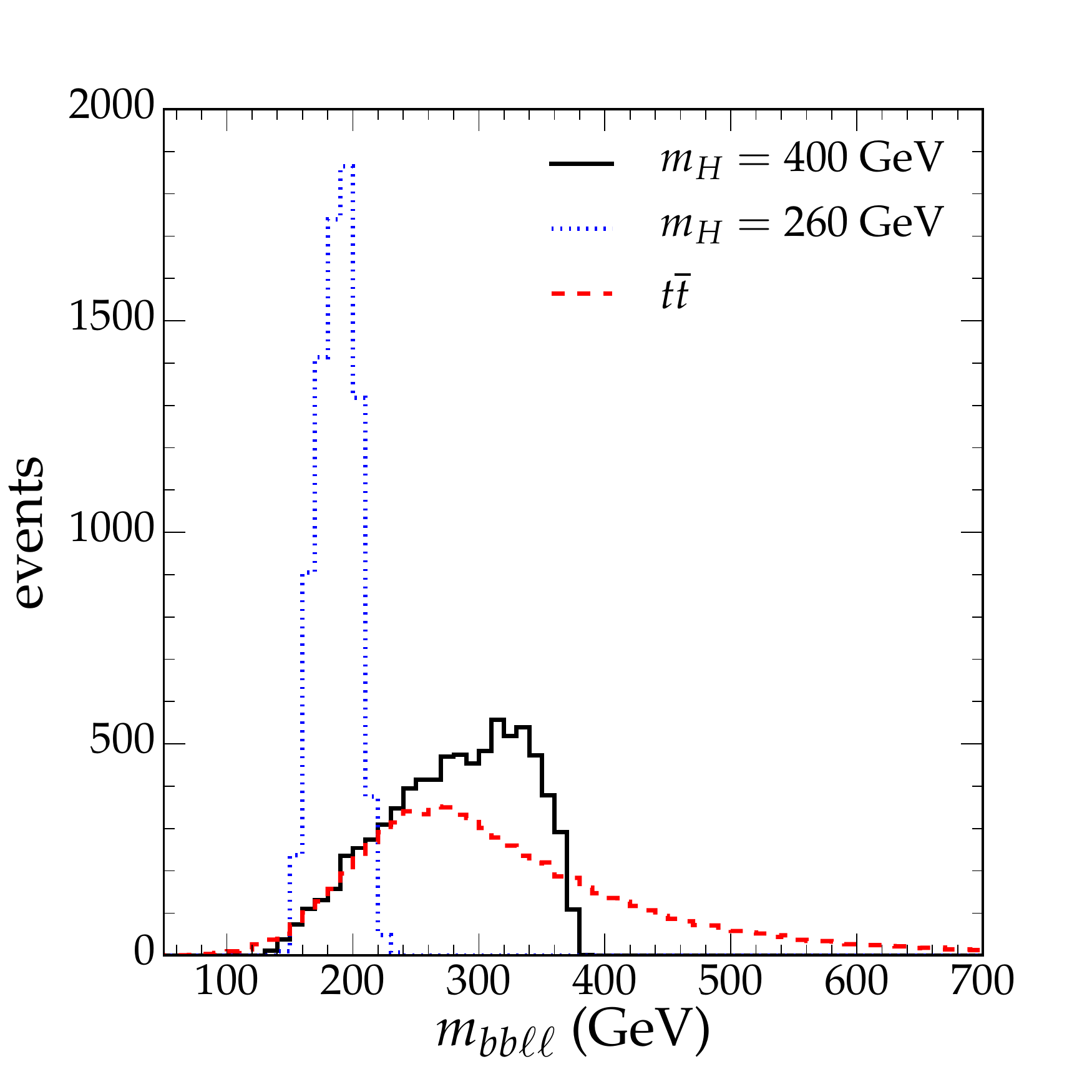}
    \includegraphics[width=0.48\textwidth]{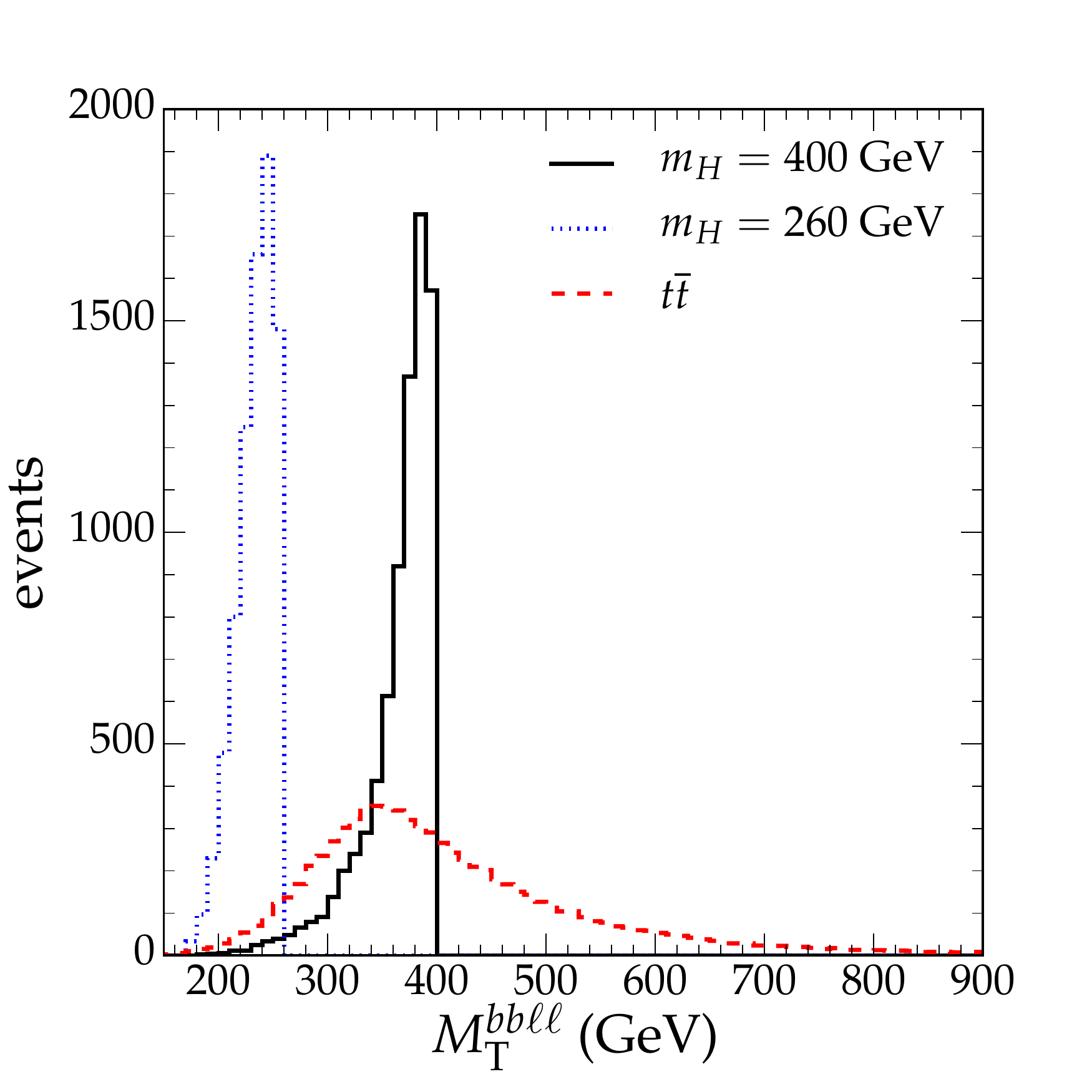}
  \end{center}
  \caption{Normalized distributions for (Left panel) $m_{bb\ell\ell}$
    and (Right panel) $M_{\rm T}^{bb\ell\ell}$ distributions for $m_H
    = 260$ and 400 GeV and the $t\bar{t}$ backgrounds using
    parton-level data.}
\label{fig:m_bbll}
\end{figure}
See the left panel of Figure~\ref{fig:m_bbll}. Since the lower bound
is fixed as $m_{b\bar{b}} = m_h = 125$ GeV in the signal events, only
upper cut on $m_{bb\ell\ell}$ variable can be applied.
For the benchmark point with $m_H = 400$ (260) GeV, we set the cut as
$m_{bb\ell\ell} < 395$ (200) GeV. This cut becomes important in the case
of a heavy Higgs with lower mass value like in the case of $m_H = 260$
GeV since it is capable of taking more stronger cut value.
The other useful kinematic variable is the transverse mass of the full
system defined as
\begin{align}
  \left ( M_{\rm T}^{bb\ell\ell} \right )^2 = m_{bb\ell\ell}^2 +
  2 \left (\sqrt{|\mathbf{p}_{\rm T}^{bb\ell\ell}|^2 +
    m_{bb\ell\ell}^2}|\slashed{\mathbf{p}}_{\rm T}| - \mathbf{p}_{\rm
    T}^{bb\ell\ell} \cdot \slashed{\mathbf{p}}_{\rm T} \right ),
\end{align}
where $\mathbf{p}_{\rm T}^{bb\ell\ell} \equiv \mathbf{p}_{\rm T}^b +
\mathbf{p}_{\rm T}^{\bar{b}} + \mathbf{p}_{\rm T}^\ell +
\mathbf{q}_{\rm T}^\ell$ is the total visible transverse
momentum. Here, the unknown $m_{\nu\nu}$ is
ignored. When all the visible particles are on the
transverse plane and the neutrino momentum vectors are collinear so
that the $m_{\nu\nu}$ is vanishing, the transverse mass is equivalent
to the invariant mass of the full system, \ie, $m_H$. It means that
the transverse mass is bounded from above by $m_H$ as can be seen in the
right panel of Figure~\ref{fig:m_bbll}. In real situation, the endpoint
of the distribution is often smeared by the backgrounds and/or poor
reconstruction efficiency of the final-state objects. Still, since the
peak position is near the endpoint, it can provide an lower bound on
$m_H$. On the other hand, the transverse mass has some
correlation with the MAOS invariant mass as discussed
in~\cite{Choi:2010dw}. They select the similar types of events in
the phase space, and the efficiency is comparable to each other. We
employ both two variables to suppress backgrounds and define the
signal region.

\begin{table}[bt!]
  \vspace{1pt}
  \begin{center}
    \begin{tabular}{ c | c | c ccc cc | c}
      \hline&&&&&&&&\\[-2mm]
      Selection cuts &
      $H \to hh$ & $t\bar{t}$ & GGF $h$ &
      $h t\bar{t}$ & $hh$ & DY & $VV$ & $\hat\sigma_{3000}$\\[2mm]
      \hline&&&&&&&&\\[-2mm]
      Basic selection &
      0.54 & 3560.36 & 0.15 &
      0.072 & 0.024 & 272.41 & 0.90 & 0.48 \\[2mm]
      $\Delta\phi_{\ell\ell}$, $\Delta R_{\ell\ell}$,
      $p_{\rm T}^{\ell\ell}$ &
      0.40 & 562.02 & 0.11 &
      0.015 & 0.019 & 33.56 & 0.047 & 0.90 \\[2mm]
      $m_{\ell\ell}$, $M_{\rm T2}^{\ell\ell}$ &
      0.36 & 314.95 & 0.097 &
      0.009 & 0.017 & 11.20 & 0.0 & 1.1 \\[2mm]
      $m_{h}^{\rm maos}$, $M_{\rm T}^{\ell\ell}$ &
      0.33 & 237.96 & 0.097 &
      0.007 & 0.015 & 11.20 & -- & 1.2 \\[2mm]
      $\Delta R_{bb}$, $p_{\rm T}^{bb}$ &
      0.23 & 73.03 & 0.008 &
      0.002 & 0.012 & 3.73 & -- & 1.4 \\[2mm]
      $m_{bb}$ &
      0.14 & 16.24 & 0.0 &
      $\simeq 0.0$ & 0.007 & 0.0 & -- & 1.9 \\[2mm]
      $\Delta \phi_{bb,\,\ell\ell}$, $m_{bb\ell\ell}$ &
      0.13 & 11.99 & -- &
      -- & 0.005 & -- & -- & 2.1 \\[2mm]
      $M_{\rm T2}^{bb\ell\ell}$ &
      0.059 & 1.31 & -- &
      -- & 0.004 & -- & -- & 2.8 \\[2mm]
      \hline&&&&&&&&\\[-2mm]
      Signal region &
      0.048 & 0.70 & -- &
      -- & $\simeq 0.0$ & -- & -- & 3.1 \\[2mm]
      \hline
    \end{tabular}
  \end{center}
  \caption{
    Cut flow of signals for $m_H = 400$ GeV and the main backgrounds
    in fb. See the text for
    detailed description of the event selection cuts applied. $VV$
    denotes the di-boson processes ($V = W,\,Z$).
    $\hat\sigma_{3000}$ is the signal significance calculated with a
    Poisson probability at 3000 fb$^{-1}$ integrated luminosity.
    The signal region is defined by $345~{\rm GeV} < M_{\rm
      T}^{bb\ell\ell} < 425$ GeV and $350~{\rm GeV} < m_H^{\rm maos} <
    430$ GeV.}
  \label{tab:cut_flow_400}
\end{table}
\begin{table}[bt!]
  \begin{center}
    \begin{tabular}{ c | c | c ccc cc | c}
      \hline&&&&&&&&\\[-2mm]
      Selection cuts &
      $H \to hh$ & $t\bar{t}$ & GGF $h$ &
      $h t\bar{t}$ & $hh$ & DY & $VV$ & $\hat\sigma_{3000}$\\[2mm]
      \hline&&&&&&&&\\[-2mm]
      Basic selection &
      0.48 & 3560.36 & 0.15 &
      0.072 & 0.024 & 272.41 & 0.90 & 0.43 \\[2mm]
      $\Delta\phi_{\ell\ell}$, $\Delta R_{\ell\ell}$,
      $p_{\rm T}^{\ell\ell}$ &
      0.28 & 818.01 & 0.15 &
      0.020 & 0.022 & 48.51 & 0.095 & 0.70 \\[2mm]
      $m_{\ell\ell}$, $M_{\rm T2}^{\ell\ell}$ &
      0.21 & 206.23 & 0.11 &
      0.006 & 0.007 & 0.0 & 0.0 & 0.80 \\[2mm]
      $m_{h}^{\rm maos}$, $M_{\rm T}^{\ell\ell}$ &
      0.19 & 140.69 & 0.08 &
      0.004 & 0.005 & -- & -- & 0.88 \\[2mm]
      $\Delta R_{bb}$, $m_{bb}$ &
      0.104 & 6.65 & 0.008 &
      $\simeq 0.0$ & $\simeq 0.0$ & -- & -- & 2.21 \\[2mm]
      $m_{bb\ell\ell}$ &
      0.009 & 3.03 & 0.008  &
      -- & -- & -- & -- & 2.82 \\[2mm]
      $M_{\rm T2}^{bb\ell\ell}$ &
      0.083 & 2.29 & 0.0 &
      -- & -- & -- & -- & 2.99 \\[2mm]
      \hline&&&&&&&&\\[-2mm]
      Signal region &
      0.083 & 2.19 & -- &
      -- & -- & -- & -- & 3.06 \\[2mm]
      \hline
    \end{tabular}
  \end{center}
  \caption{
    Cut flow of signals for $m_H = 260$ GeV and the main backgrounds
    in fb. See the text for
    detailed description of the event selection cuts applied.
    $\hat\sigma_{3000}$ is the signal significance calculated with a
    Poisson probability at 3000 fb$^{-1}$ integrated luminosity.
    The signal region is defined by $180~{\rm GeV} < M_{\rm
      T}^{bb\ell\ell} < 265$ GeV and $185~{\rm GeV} < m_H^{\rm maos} <
    305$ GeV.}
\label{tab:cut_flow_260}
\end{table}

Combining all the cuts discussed so far, we examine their effects on
the signal and the backgrounds by investigating how the cross sections
are changing by applying event selection cuts. See
Tables~\ref{tab:cut_flow_400} and \ref{tab:cut_flow_260} for $m_H =
400$ and 260 GeV, respectively. The unlisted backgrounds turned
out to be almost negligible after applying the initial cuts.
In summary, although the
production cross section for $m_H = 400$ GeV is smaller than that of
$m_H = 260$ GeV, the signal can be distinguished by several angular
cut variables as well as the cut on $M_{\rm
  T2}^{bb\ell\ell}$. We have found a set of kinematic
variables useful for the search.
Eventually, the scenarios with a relatively
lighter singlet-like Higgs boson are quite difficult to probe by using
the kinematic event variables. In this case, one can still attempt to
combine the search results from the other channels like $bb\tau\tau$ and
$bbZZ$, which have the next-to-subleading branching fractions, or a
multivariate analysis like performed
in~\cite{Papaefstathiou:2012qe}. If $m_H$ is much larger than 400 GeV,
it is expected that the boosted Higgs technique approach is more
promising.

\begin{figure}[ht]
  \begin{center}
    \includegraphics[scale=0.53]{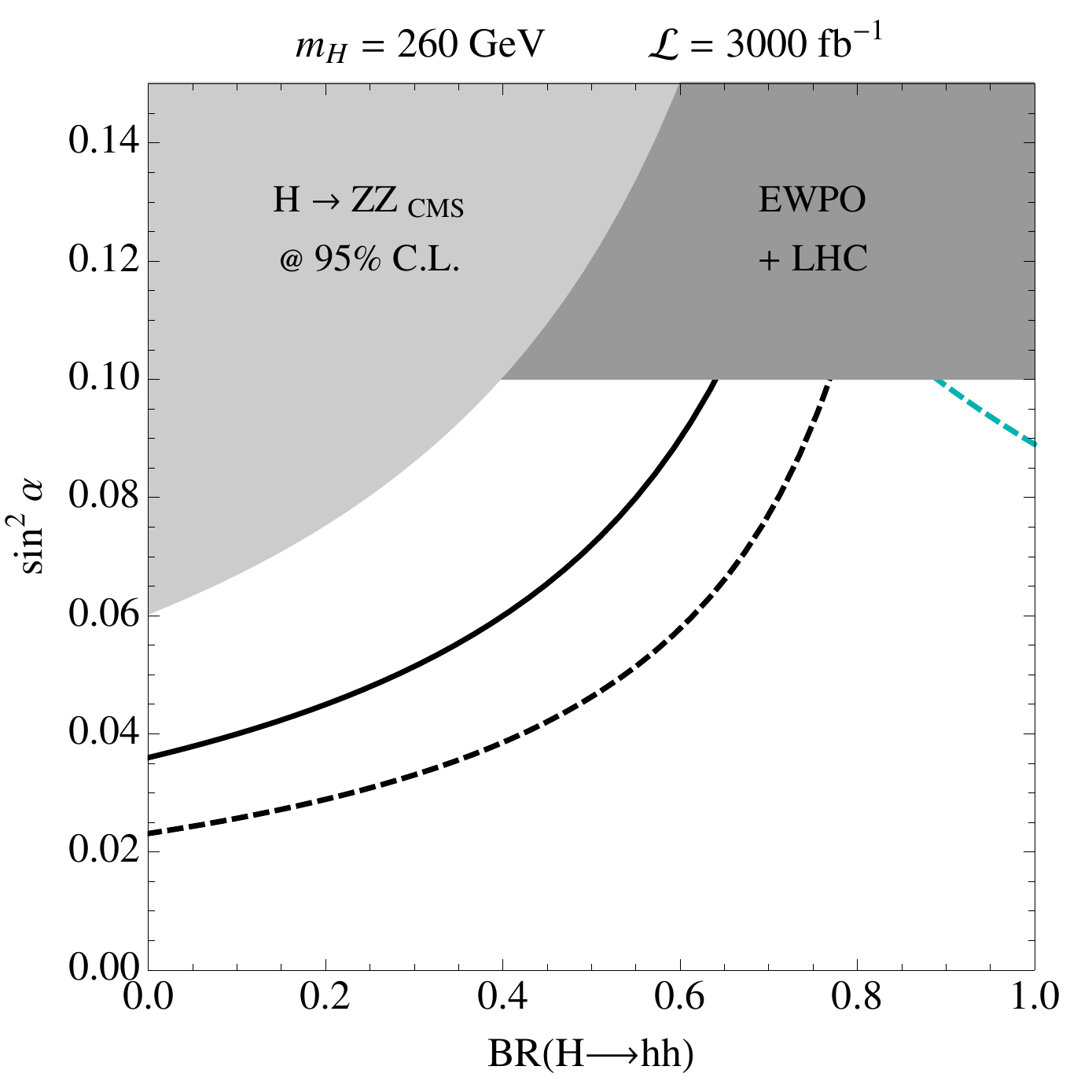}
    \includegraphics[scale=0.53]{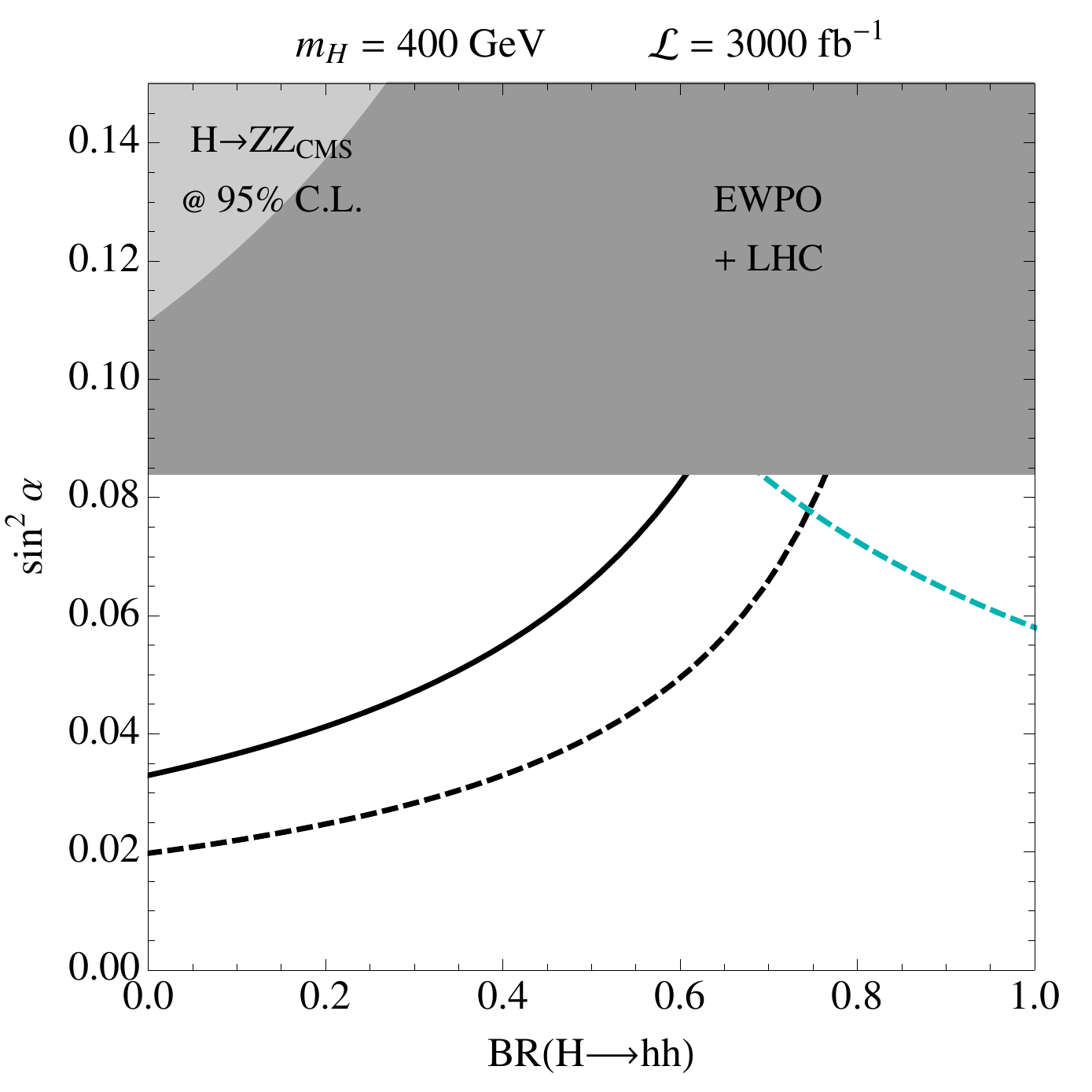}
  \end{center}
  \caption{The dashed line delimits the 3$\sigma$ significance
    region in the the $\sin^2\alpha - \BR(H \to hh)$ plane for
    the $H \to hh \to b \bar{b} W W^\ast \to 2b + 2 \ell  + 2
    \nu$ $(\ell = e,\,\mu)$ process for the integrated luminosity
    of 3000~fb$^{-1}$.
    The solid (dashed) black curve corresponds to the 5$\sigma$
    (3$\sigma$) for the $H \to ZZ  \to  4\ell / 2\ell 2\tau$ channels.
    Dark grey shaded region is the 95\% C.L. CMS exclusion
    bounds and the light grey region is the one for EWPO $+$
    LHC.}
  \label{fig:detectablity}
\end{figure}

Up to now, we have assumed that $\BR(H \to hh) \sim 1$. This can be
fulfilled in a large $(\lambda_{22}, \, \lambda_{03}, \,
\lambda_{03})$ parameter-space region. We now relax this condition and
suppose that  the SM Higgs-like decays  originated by the mixing are
non-negligible. In this case, for a given $m_H$ value, we can evaluate
bounds on the mixing using the ATLAS and CMS data on heavy Higgs
searches~\cite{Chatrchyan:2013yoa}, as shown in
Figure~\ref{fig:CMSBounds}.
The most stringent exclusion limit comes from the CMS search~\cite{CMS:2013pea,CMS:xwa}.
This search is focused on the combination of the $4\ell$/$2\ell 2\tau$
final states in the $H\to ZZ$ channel assuming that the heavy Higgs only decays into SM
particles, \ie, $\BR (H \to hh)$ is vanishing. The maximal mixing
angle allowed by this search
for $m_H =260$~GeV is $\sin^2 \alpha < 0.06$ (95\% C.L.), while for
$m_H = 400$~GeV it is $\sin^2 \alpha < 0.11$ (95\% C.L.).
If $\BR(H \to hh)$ is non-vanishing, the latter constraints
become weaker.
The excluded $\sin^2 \alpha$ values for given $\BR (H \to
hh)$ are represented in the light gray region in
Figure~\ref{fig:detectablity} for both $m_H =260$~GeV and $m_H =
400$~GeV.
On the other hand, the constraints
imposed by the EWPO and the LHC, shown as dark gray region in
Figure~\ref{fig:detectablity}, are independent of $\BR (H \to hh)$.
This is because they come from the modification of the couplings,
parameterized by the mixing angle $\alpha$, while the ones derived
from the heavy Higgs searches depend directly on the value of $\BR (H
\to hh)$.
One can also see the interplay between direct and indirect
constraints in Figure~\ref{fig:detectablity}. For $m_H =260$~GeV,
the direct search result on $H\to ZZ$ imposes the
most stringent bound, up to $\BR (H \to hh)\sim 0.4$. For larger
values of $\BR (H \to hh)$, the LHC $+$ EWPO limits are the most
important since the direct search limit weakens. For $m_H
=400$~GeV the direct limit is not as stringent as the indirect
ones, which impose an upper bound of $\sin^2\alpha < 0.084$,
independently of the $\BR (H \to hh)$ value.

We can use the discovery reach of the 14 TeV
LHC~\cite{Ball:2007zza} for the Higgs boson search using the decay
channel $H \to ZZ  \to  4\ell$ in order to estimate the
detectability of the two $m_H$ values as a function of the mixing and
the $\BR (H \to hh)$.\footnote{We assume that $\BR (H \to hh) + \BR (H
  \to \text{SM particles}) = 1$.}
In Figure~\ref{fig:detectablity}, we show the
$3\sigma$ and  $5\sigma$ significance lines  for this channel for the
integrated luminosity of 3000~fb$^{-1}$. These lines show that for low
values of $\BR (H \to hh)$ this search is able to resolve a large portion
of the mixing angle values, leaving a small window of possible values.
The sensitivity of this channel begin to decrease for $\BR (H \to hh)>0.6$, just in
the region where the double Higgs production, in particular the channel above mentioned, becomes relevant.
In Figure~\ref{fig:detectablity}, we have included the $3\sigma$ equivalent line for the $H \to hh \to b \bar{b}
W W^\ast$ channel. It is important to note that both channels are complementary
since they are very dependent on the value of $\BR (H \to hh)$.
As a remark the ATLAS
collaboration has performed a search of heavy Higgses using the
channel $H\to hh \to b\bar{b}\gamma\gamma$~\cite{ATLAS:bbgg}. The
results are not shown because the exclusion limit is well above the
ones appeared in Figure~\ref{fig:detectablity}.

\section{Comments on Dark Matter}
\label{sec:darkmatter}

Given that the new scalar is unstable, it does not solve the dark
matter problem. Nevertheless, it can play a relevant role by
providing a portal to DM.
In this section we explore this possibility. The DM mass and its coupling to the new scalar will be restricted by
requiring a DM relic density in agreement  with the experimental
value. We analyse the compatibility between this condition and
the requirement of a sizable $H \to hh$ branching ratio, as
assumed in the previous section.

Let us consider an extra singlet neutral Dirac fermion transforming
under a $Z_2$ symmetry. There is a unique even renormalizable
interaction term, so the Lagrangian gets enlarged by
\begin{equation}
 \bar{\psi}(i\slashed{\partial}-m_0)\psi +
 \lambda_{\psi}S\bar{\psi}\psi .
\end{equation}
The singlet fermion is stable due to the $Z_2$ parity and is then a
potential, WIMP-like,  DM candidate.

\subsection{Relic density}

\begin{figure}
\begin{center}
\includegraphics[scale=0.52]{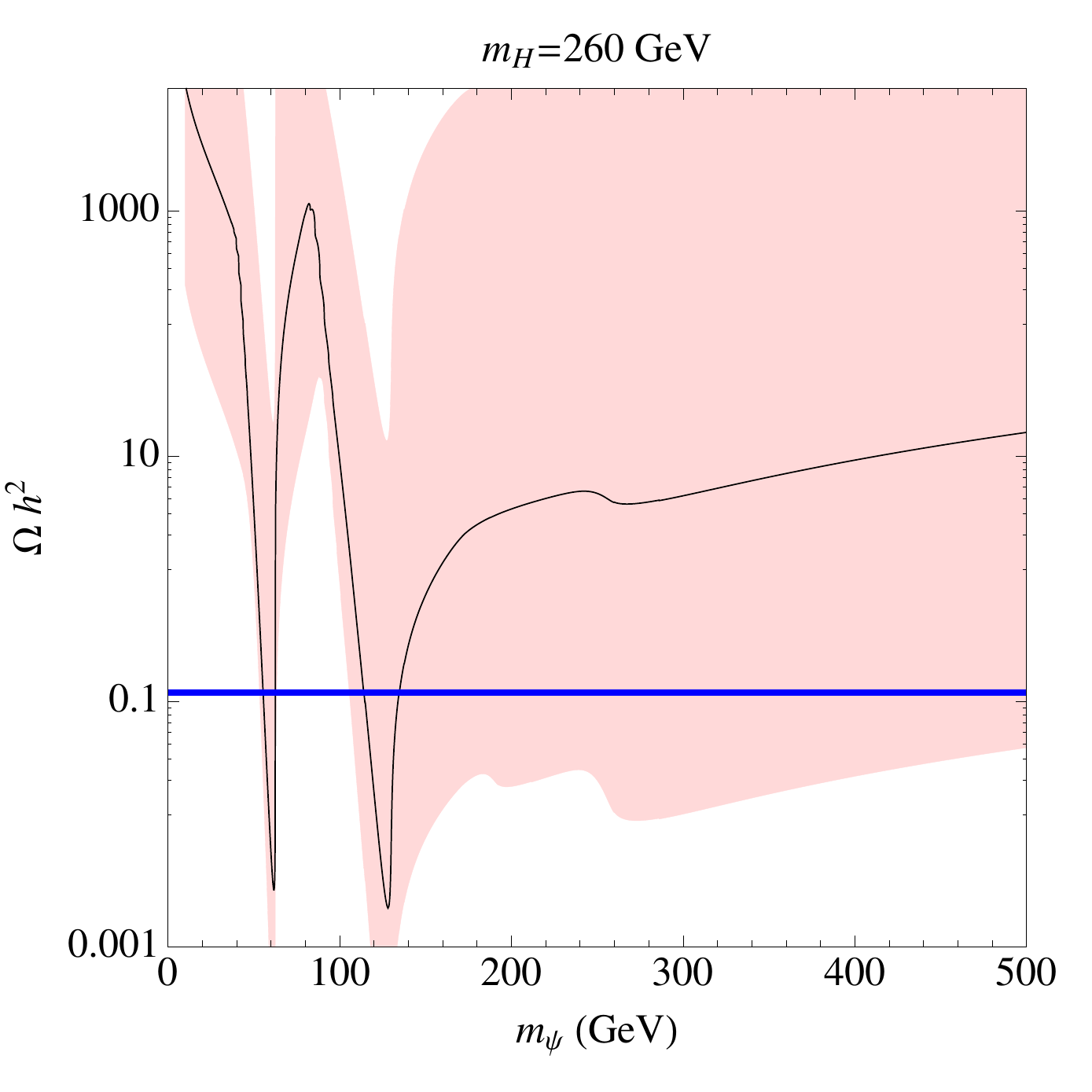}
\includegraphics[scale=0.52]{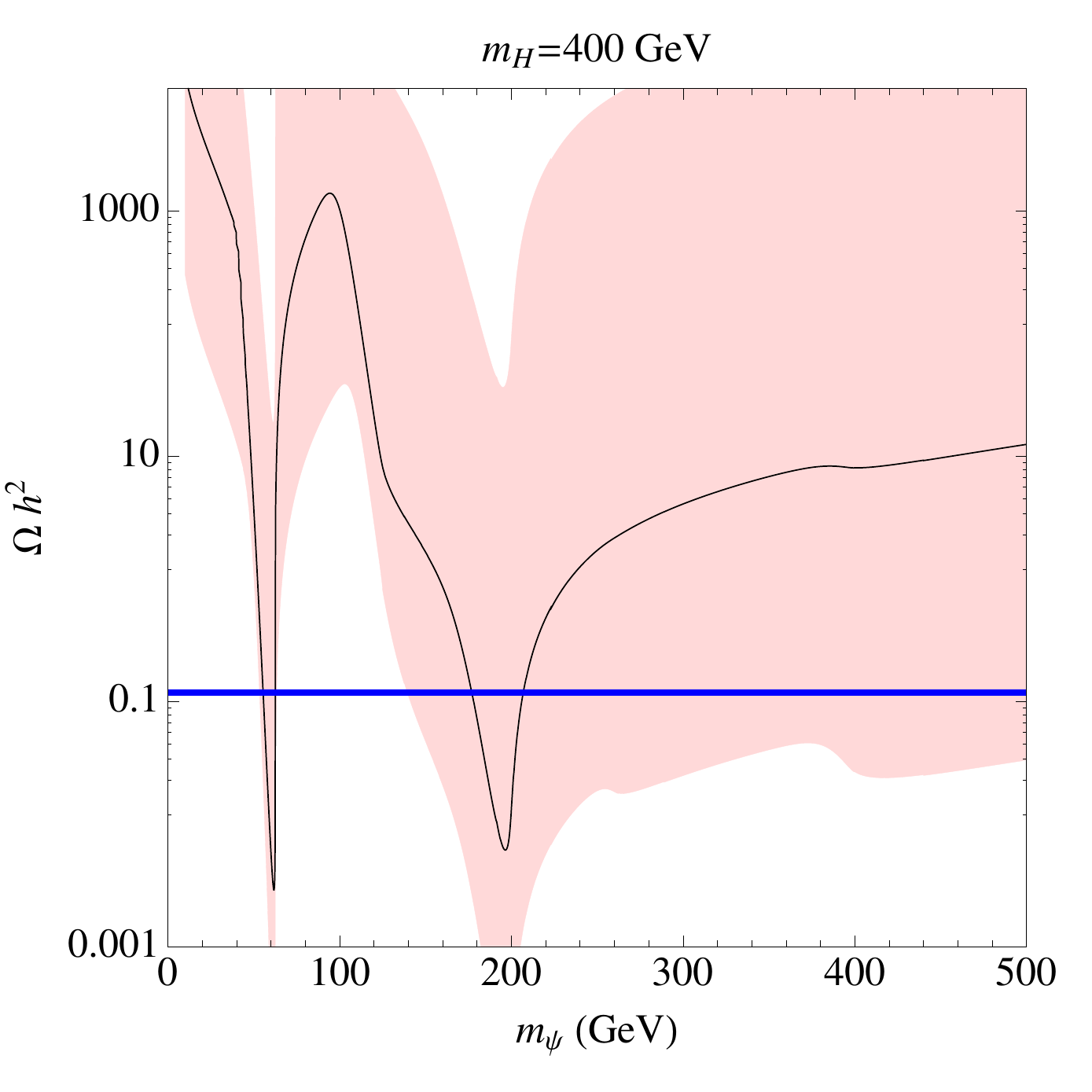}
\end{center}
\caption{
DM relic density as a function of the WIMP mass, $m_\psi$, for
   different values of $\lambda_\psi$. See the text for detailed
   description.
 }
\label{fig:omega}
\end{figure}

We have implemented the model in~\textsc{CalcHEP}~\cite{Belyaev:2012qa}
and used the \textsc{micrOMEGAs 2.4} package~\cite{Belanger:2010gh} to
evaluate the DM relic density for the two benchmark points studied in
the previous section.
The results are displayed in Figure~\ref{fig:omega}, where we show the
DM relic density as a function of the WIMP mass, $m_\psi$, for
   different values of $\lambda_\psi$. The light red region
   corresponds to $\lambda_\psi$ values varying from 0.001 to 1. The black
   solid line represents the relic density for 
   $\lambda_\psi = 0.1$. The blue band is bounded by the
   allowed experimental relic density value given by Planck~\cite{Ade:2013zuv}:
\begin{equation}
0.1134<\Omega h^2<0.1258  \; \; ( 95\% \,  \text{C.L})
\end{equation}
Note that the correct relic density can be achieved in two regions.
The first one is characterized by a DM mass close to $m_h/2$,
providing an enhancement of the DM annihilation cross section due to
the resonance effect. When kinematically allowed, the Higgs decay into
a $\psi$ pair becomes to be dominant.
As the LHC  constrains the Higgs invisible width, that is mainly given by
\begin{equation}
  \Gamma ( h \to \psi \bar{\psi})   =   \frac{| \lambda_\psi \sin \alpha |^2}
  {16\pi} m_h \left(1-\frac{4m_{\psi}^2}{m_{h}^2}  \right)^{3/2},
\end{equation}
this small $m_\psi$ window gets reduced ($\sim 1$~GeV).

There is a much wider parameter region where the enough amount of
   DM annihilation can be triggered by the heavy Higgs. Around and above 
   the region of the heavy Higgs resonance, \ie,
   $2m_\psi \gtrsim m_H$, the other annihilation channels such as $\psi
   \bar \psi \to h H$ and $\psi \bar \psi \to HH$ are open, thus
   making the DM annihilation sufficient to attain the correct relic
   density. For $2 m_\psi < m_H $, the
$H \to \psi\psi$ decay process will contribute to the decay width
   of the heavy Higgs boson, reducing the $\BR(H \to
hh )$ ratio and then decreasing the resonant double Higgs
production. This could affect the analysis done in the previous sector
by reducing the statistical significance of the signal. However in the
region where $2 m_\psi > m_H$ the results would remain
unaffected. For this reason we should incorporate the constraints 
from the direct detection experiments in order to
know which DM regions are favoured. A similar study was done
in~\cite{Li:2014wia}, that agrees with our analysis.

\subsection{Direct Detection}

\begin{figure}
\begin{center}
\includegraphics[scale=0.36]{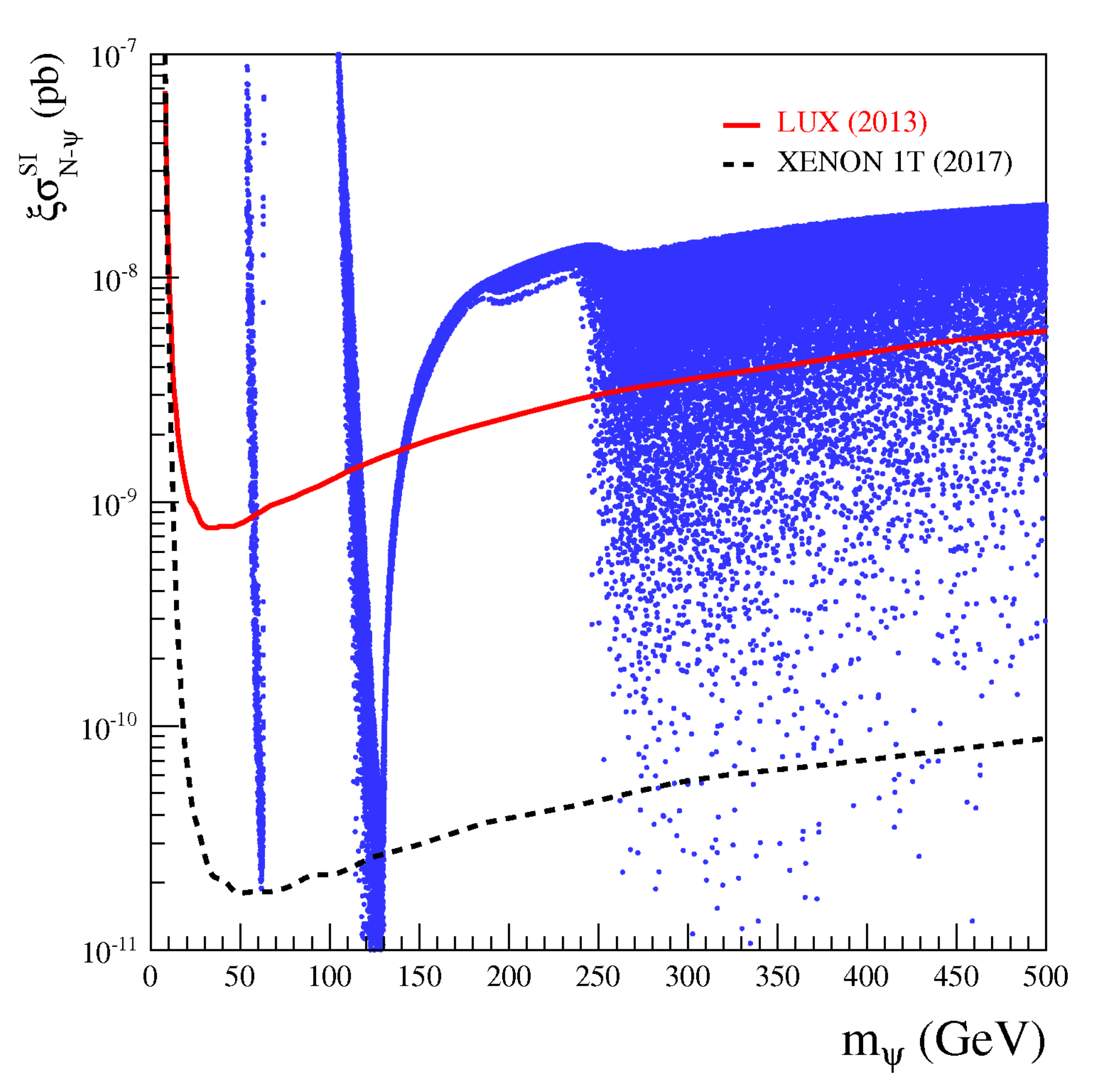}
\includegraphics[scale=0.36]{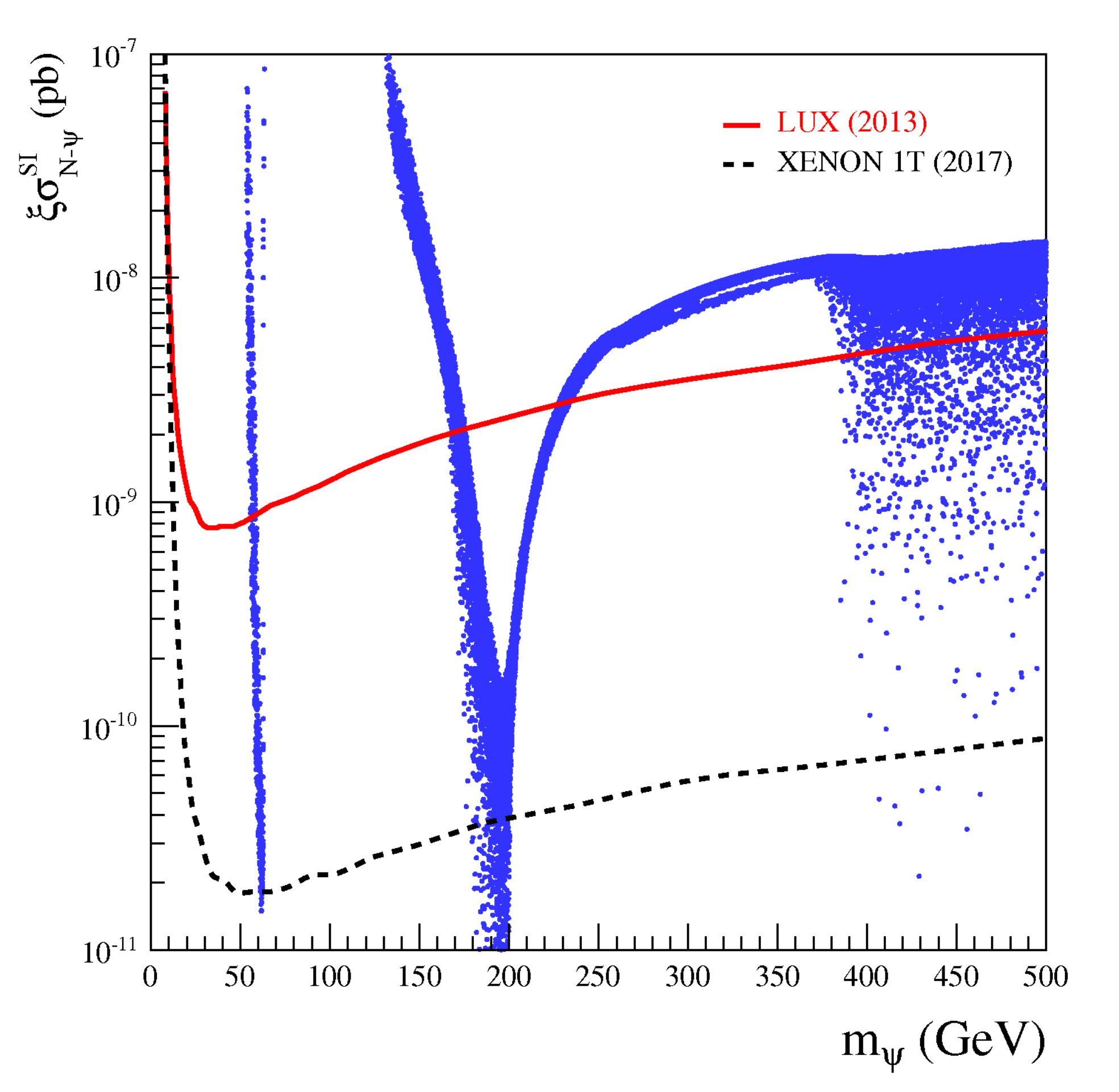}
\end{center}
\caption{Spin independent cross section as function of the DM mass for
  the two different scenarios with $m_H=260$ GeV (left) and $m_H=400$
  GeV (right). The red line represents the bounds from
  LUX~\cite{Akerib:2013tjd} while the black dashed line corresponds to
  the future prospects of XENON1T~\cite{Aprile:2012zx}.}
\label{fig:sics}
\end{figure}

Direct detection experiments search for DM by means of its elastic
scattering off nuclei. In the absence of a positive signal, present
search results translate into bounds on the WIMP-nuclei cross section for a
given WIMP mass. As the elastic scattering is produced at low
momentum we can write the interaction as an effective operator. In
our case, it is induced by $t$-channel exchange of the Higgses and is
given by:
\begin{equation}
  \mathcal{L}_{\rm eff} \supset\alpha_{q_i}\bar{\psi}\psi \bar{q}_iq_i,
\end{equation}
with~\cite{Kim:2008pp}
\begin{equation}
\frac{\alpha_q}{m_q}=\frac{\lambda_\psi\cos\alpha\sin\alpha}{v}\left(\frac{1}{m_h^2}-\frac{1}{m_H^2}\right),
\end{equation}
The spin-independent elastic scattering cross section can be written
as\footnote{See, for example, ref.~\cite{Cerdeno:2010jj}}
\begin{equation}
\sigma_{\psi p}^{\rm{SI}}=\frac{1}{\pi}\frac{m_p^2}{(m_p+m_{\psi})^2}f_p^2,
\end{equation}
where $m_p$ is the proton mass and $f_p$ is defined as
\begin{equation}
\frac{f_p}{m_p}=\sum_{q_i=u,d,s}f_{Tq_i}^p\frac{\alpha_{q_i}}{m_{q_i}}+\frac{2}{27}f_{TG}^{p}\sum_{q_i=c,b,t}\frac{\alpha_{q_i}}{m_{q_i}}
\end{equation}
where the quantities $f_{Tq_i}$ represent the contributions of the light quarks to the mass of the proton.
The full expressions for the spin-independent cross section can be
found in refs.~\cite{Kim:2008pp, Fairbairn:2013uta}.
In Figure~\ref{fig:sics} the normalized spin-independent cross section
is plotted as a function of the DM candidate mass for the two
benchmark points.
This normalized cross section, $\xi \sigma_{\psi N}^{\rm SI}$, is the
product of the spin-independent cross section and the factor $\xi$
defined as $\xi\equiv \min\{1,\, \Omega_{\psi}h^2/0.1226\}$.
This factor accounts for situations where $\psi$ provides only a
fraction of the total amount of dark matter.
In Figure~\ref{fig:sics} a scan over the mass and the
$\lambda_\psi$ parameters has been done. Only the points with a relic
density equal or less than that from Planck are showed. The bounds
imposed by LUX~\cite{Akerib:2013tjd} are included as well as  future
prospect from XENON 1T~\cite{Aprile:2012zx}.

For the light DM candidate it is difficult to have the
correct relic density and avoid the bound imposed by LUX at the same
time.
These conditions are compatible in a small region close to half of the
mass of the Higgses so a resonant peak is present.
However, this means that the decays $h\to \psi\bar{\psi}$ and
$H\to \psi\bar{\psi}$ are dominant, so the model could be excluded by
the LHC or would spoil the results obtained in the collider
analysis. Nevertheless we can find a
region with relatively high masses of the DM candidate that fulfills
both relic density and spin-independent cross section and is
placed above the resonance produced by the heavy Higgs.
In fact, the allowed area is induced by the opening of the  $\psi\bar{\psi}\to H H$ annihilation channel, 
making the cross section more effective.

To summarize,  our analysis implies that there is a region where DM requirements are fulfilled and is located  above the heavy Higgs mass so the constraints from the LHC and the results obtained in the collider analysis are not
affected. Furthermore we can see that in the next years direct
detection experiments such as XENON1T are sensitive to a large amount
of the parameter space of this model, leading to the possibility of
probing it.

\section{Conclusions}
\label{sec:conclusion}

In the coming years, the LHC will further explore  the properties of the Higgs
boson by looking for possible deviation from the SM
predictions~\cite{Dawson:2013bba}. In particular, after the
high-luminosity upgrade LHC is expected to deliver 3000~fb$^{-1}$ at 14
TeV~\cite{Peskin:2013xra}. This would allow to measure the $\gamma
\gamma$, $WW$, $ZZ$, $b\bar{b}$, and $ \tau^+\tau^-$ Higgs couplings
within a 2 -- 8\% error~\cite{Dawson:2013bba, Peskin:2013xra}.
Meanwhile, the singlet-extended model is the simplest extension of the
SM scalar sector. It predicts a universal deficit in the Higgs boson
couplings to the SM fermions and gauge bosons caused by the mixing
between the two neutral scalar states.
Alternatively, a new contribution to the invisible Higgs width would imply the
reduction of the visible Higgs decays, that can also be interpreted as
a generic Higgs coupling deficit. The direct production and detection
of the new Higgs would certainly elucidate this point. Since the
relevant cross section depends on the mass and the mixing, we have
first reviewed the present experimental bounds on these two
parameters. Concerning the constraints by EWPO, we have improved
previous analysis by using the full set of electroweak observables
instead of the oblique parameters $(S,\,T)$ since the last ones only
provide an accurate descriptions of the heavy Higgs effects in the
$m_H \sim m_h$ region.

In order to illustrate the detection of the direct heavy Higgs
production, we have chosen
two benchmark points compatible with present bounds, in particular,
the LHC Higgs data and the EWPO.
We have studied the resonant SM Higgs boson pair production in the
$hh\to b\bar{b} \; WW \to b \bar{b} \ell^+ \nu \ell^-
\bar{\nu}$ decay channel. The main background to the signal is the
di-leptonic $t \bar{t}$ process. Besides some basic selection cuts, we
have applied $M_{\rm T2}$ cuts for the $2\ell + \slashed{E}_{\rm T}$
or $2 b + 2\ell + \slashed{E}_{\rm T}$
systems to optimise the signal significance. Using the di-leptonic
channel alone, a significance $\sim 3\sigma$ for 3000~fb$^{-1}$ can be
achieved at the 14~TeV LHC for $m_H = 400$~GeV if the
mixing is close to its present limit  and  $\BR(H \to hh) \approx
1$. A lower branching ratio or a smaller mixing angle would require
combining various $hh$ decay channels. The complementarity between $H
\to hh$ and $H \to ZZ$ channels is studied for arbitrary $\BR(H \to
hh)$ values.

We have also checked that it is possible to extend the model by
including a DM candidate. The next generation of direct detection
experiments will be capable of probing a large amount of the parameter
space of the model.

{\em Note added:} After completion of this work, some similar
results have been presented in~\cite{Robens:2015gla}.

\section*{Acknowledgments}
VML is very grateful to the ZFITTER authors, specially to Tord Riemann, for useful and valuable explanations about ZFITTER program.
JMM and VML are partially supported by the grants FPA2010-17747,
FPA2012-34694,  FPA2013-44773-P and from the Spanish MINECO, Consolider-Ingenio CPAN
CSD2007-00042 and MULTIDARK CSD2009-00064 as well as  the Centro de
Excelencia Severo Ochoa Program under Grant No. SEV-2012-0249.
VML also thanks support by the ERC Advanced Grant SPLE under contract
ERC-2012-ADG-20120216-320421.  The work of CBP was partially supported by the CERN-Korea  fellowship through National Research Foundation of Korea.

\section*{Appendix}
\appendix

\section{\boldmath Higgs $M_{\rm T2}$ in di-leptonic $WW$ process}
\label{sec:m_T2_Higgs}

We here consider the maximum of $M_{\rm T2}$ in the di-leptonic
decay mode of the $h \to WW^{(\ast)}$ process,
\begin{align}
  p p \to h + j \to W W^{(\ast)} + j
  \to \ell (p) \nu (k) + \ell (q) \nu (l) + j(u),
  \label{eq:SM_Higgs_decay}
\end{align}
where $j$ denotes the initial state radiation, typically jets in the
final state. The transverse components of the total momentum should be
conserved and therefore
\begin{align}
  \mathbf{p}_{\rm T} + \mathbf{k}_{\rm T} + \mathbf{q}_{\rm T} +
  \mathbf{l}_{\rm T} + \mathbf{u}_{\rm T} = \mathbf{0}.
\end{align}
Since the visible particle in each decay chain, \ie,
the charged leptons are massless, the $M_{\rm T2}$ value is always
achieved in a balanced configuration in which $M_{\rm T}^{(1)} =
M_{\rm T}^{(2)}$ or
\begin{align}
  \left ( |\mathbf{p}_{\rm T}|  + |\mathbf{k}_{\rm T}^{\rm maos}|
  \right )^2
  - | \mathbf{p}_{\rm T} + \mathbf{k}_{\rm T}^{\rm maos} |^2
  &=
  \left ( |\mathbf{q}_{\rm T}|  + |\mathbf{l}_{\rm T}^{\rm maos}|
  \right )^2
  - | \mathbf{q}_{\rm T} + \mathbf{l}_{\rm T}^{\rm maos} |^2,
  \nonumber\\
  \mathbf{k}_{\rm T}^{\rm maos} + \mathbf{l}_{\rm T}^{\rm maos}
  &= - \mathbf{p}_{\rm T} - \mathbf{q}_{\rm T} - \mathbf{u}_{\rm T},
  \label{eq:balanced}
\end{align}
where $\mathbf{k}_{\rm T}^{\rm maos}$ and $\mathbf{l}_{\rm T}^{\rm
  maos}$ stand for the $M_{\rm T2}$ solution for the invisible
transverse momenta.
The above equations are satisfied when $\mathbf{k}_{\rm T}^{\rm maos} =
- \mathbf{q}_{\rm T}$ and $\mathbf{l}_{\rm T}^{\rm maos} =
- \mathbf{p}_{\rm T}$ for vanishing $\mathbf{u}_{\rm T}$.
On the other hand, if $\mathbf{u}_{\rm T}$ is sizable, the solution
can be redefined as
\begin{align}
  \mathbf{k}_{\rm T}^{\rm maos} &= - \mathbf{q}_{\rm T}
  - \frac{\mathbf{u}_{\rm T}}{2} - \delta \mathbf{u}_{\rm T}
  \nonumber\\
  \mathbf{l}_{\rm T}^{\rm maos} &= - \mathbf{p}_{\rm T}
  - \frac{\mathbf{u}_{\rm T}}{2} + \delta \mathbf{u}_{\rm T} .
\end{align}
where $\delta \mathbf{u}_{\rm T}$ is a function parameterizing the
transverse boost effect of the solution by $\mathbf{u}_{\rm T}$ while
preserving conditions (\ref{eq:balanced}) and $\delta \mathbf{u}_{\rm
  T}(\mathbf{u}_{\rm T} = \mathbf{0}) = 0$.\footnote{
  See~\cite{Lester:2011nj} for the dedicated discussion of the
  $M_{\rm T2}$ solution in the case of fully massless visible and
  invisible particles.
}
This solution is generically different from the true invisible momenta,
\ie, $\mathbf{k}_{\rm T}^{\rm maos} \neq \mathbf{k}_{\rm T}$ and
$\mathbf{l}_{\rm T}^{\rm maos} \neq \mathbf{l}_{\rm T}$, however by
construction, the sum of each component must be equal,
\begin{align}
  \mathbf{k}_{\rm T}^{\rm maos} + \mathbf{l}_{\rm T}^{\rm maos} =
  \mathbf{k}_{\rm T} + \mathbf{l}_{\rm T}.
\end{align}

The maximum value of $M_{\rm T2}$ for given visible momenta and the
sum of invisible momenta can be deduced from a kinematic property as
given below,
\begin{align}
  M_{\rm T} \leq m_h,
  \label{eq:trans_inv}
\end{align}
where $M_{\rm T}$ and $m_h$ are the transverse and the invariant
masses for two charged leptons and two neutrinos defined as
\begin{align}
   M_{\rm T}^2 = &
  \left ( M_{\rm T}^{(1)} \right )^2 + \left ( M_{\rm T}^{(2)}
  \right )^2 + 2 \sqrt{|\mathbf{p}_{\rm T} + \mathbf{k}_{\rm T}^{\rm
      maos}|^2 + \left (M_{\rm T}^{(1)} \right)^2}
  \sqrt{|\mathbf{q}_{\rm T} + \mathbf{l}_{\rm T}^{\rm maos}|^2 + \left
      (M_{\rm T}^{(2)} \right)^2} \nonumber\\
  & - 2 \left( \mathbf{p}_{\rm T} + \mathbf{k}_{\rm T}^{\rm maos}
  \right ) \cdot \left( \mathbf{q}_{\rm T} + \mathbf{l}_{\rm T}^{\rm maos}
  \right ) , \nonumber \\
  m_h^2 = & (p + k + q + l)^2,
\end{align}
respectively. The transverse mass can be further simplified as
\begin{align}
  M_{\rm T}^2 = A^2 - |\mathbf{C}|^2 + B^2 - |\mathbf{D}|^2
  + 2 A B - 2 \mathbf{C} \cdot \mathbf{D},
\end{align}
where
\begin{equation}
  \begin{array}{ll}
  A \equiv |\mathbf{p}_{\rm T}| + |\mathbf{k}_{\rm T}^{\rm maos}|,
  &
  B \equiv |\mathbf{q}_{\rm T}| + |\mathbf{l}_{\rm T}^{\rm maos}|, \\
  \mathbf{C} \equiv \mathbf{p}_{\rm T} + \mathbf{k}_{\rm T}^{\rm maos},
  &
  \mathbf{D} \equiv \mathbf{q}_{\rm T} + \mathbf{l}_{\rm T}^{\rm maos}.
  \end{array}
\end{equation}
One can see that the equations in (\ref{eq:balanced}) is
satisfied for the kinematic configuration where
\begin{align}
  B = A, \quad \mathbf{D} = \epsilon \mathbf{C} \quad (\epsilon^2 =
  1).
  \label{eq:kinematic_conf}
\end{align}
Here, $\epsilon = -1$ corresponds to a kinematic configuration with
vanishing $\mathbf{u}_{\rm T}$, while it is non-vanishing and $\delta
\mathbf{u}_{\rm T} = \mathbf{p}_{\rm T} - \mathbf{q}_{\rm T}$ in the
case of $\epsilon = 1$.
Then, it is true that
\begin{align}
  M_{\rm T}^2 = 4A^2 - 4|\mathbf{C}|^2 -
  (\epsilon - 1) (\epsilon + 3) |\mathbf{C}|^2 \geq 4A^2 -
  4|\mathbf{C}|^2 = 4 \left (M_{\rm T}^{(1)}\right)^2.
\end{align}
for both cases. Since $M_{\rm T2} = M_{\rm T}^{(1)} = M_{\rm
  T}^{(2)}$, the relation (\ref{eq:trans_inv}) implies that
\begin{align}
  M_{\rm T2} \leq \frac{m_h}{2},
  \label{}
\end{align}
where the equality holds when eqs.~(\ref{eq:kinematic_conf}) are
satisfied. Up to now, we have not made any assumption whether the
parent $W$ boson is  on-shell. In the case that both $W$ bosons are
on-shell, $M_{\rm T2}$ is bounded from above by
$m_W$. Collectively, one can express the maximum of $M_{\rm T2}$
as
\begin{align}
  M_{\rm T2} \leq \min \left \{ m_{W}, \, \frac{m_h}{2} \right \} .
\end{align}

\bibliographystyle{JHEP}    
\bibliography{MMP}   

\end{document}